\documentclass[useAMS,usenatbib]{mnras}
\usepackage{newtxtext,newtxmath}
\usepackage[T1]{fontenc}
\usepackage{ae,aecompl}
\usepackage{float}
\usepackage{graphicx}	
\usepackage{amsmath}	
\usepackage{hyperref}
\usepackage[normalem]{ulem}
\usepackage{xcolor} 
\usepackage{appendix}
\usepackage{chngcntr}
\usepackage{siunitx}

\newcommand{\serenet}{\texttt{SERENEt}}
\newcommand{\segunet}{\texttt{SegU-Net}}
\newcommand{\recunet}{\texttt{RecU-Net}}
\newcommand{\eos}{\textit{EOS}}
\newcommand{\hi}{{\sc Hi}}
\newcommand{\hii}{{\sc Hii}}

\newcolumntype{M}[1]{>{\centering\arraybackslash}m{#1}} 

\defcitealias{Bianco2021segunet}{Paper I}
\defcitealias{Bianco2023deep2}{Paper II}

\title[21-cm \serenet{}]{Deep learning approach for identification of \hii{} regions during reionization in 21-cm observations -- III. image recovery}
\author[M. Bianco et al.]{Michele Bianco$^{1,2}$\thanks{Contact e-mail: mbianc@ethz.ch}, Sambit. K. Giri$^3$, Rohit Sharma$^{4,5}$, Tianyue Chen$^1$, Shreyam Parth Krishna$^1$,\newauthor Chris Finlay$^6$, Viraj Nistane$^6$, Philipp Denzel$^7$, Massimo De Santis$^8$, Hatem Ghorbel$^8$\\
$^1$ Laboratoire d’Astrophysique, Ecole Polytechnique Federale de Lausanne (EPFL), Observatoire de Sauverny, Versoix 1290, Switzerland \\
$^2$ Institute for Particle Physics and Astrophysics, ETH Zurich, Wolfgang-Pauli-Str 27, 8093 Zurich, Switzerland\\
$^3$ Nordita, KTH Royal Institute of Technology and Stockholm University, Hannes Alf\'vens v\"ag 12, SE-106 91 Stockholm, Sweden\\
$^4$ Institute for Data Science, FHNW University of Applied Sciences \& Arts Northwestern Switzerland, Bahnhofstrasse 6, Windisch, 5210, Switzerland \\
$^5$ Space, Planetary \& Astronomical Sciences \& Engineering (SPASE), Indian Institute of Technology, Kanpur, 208016, Uttar Pradesh, India\\
$^6$ D\'epartement de Physique Th\'eorique and Center for Astroparticle Physics, Universit\'e de Geneve, 24 quai Ernest Ansermet, 1211 Geneve 4, Switzerland \\
$^7$ Centre for Artificial Intelligence, ZHAW Zurich University of Applied Sciences, Technikumstrasse 71, 8400 Winterthur, Switzerland\\
$^8$ Haute Ecole Arc Ing\'enierie, University of Applied Sciences and Arts Western Switzerland (HES-SO), Espace de l'Europe 11, Neuchâtel, Switzerland 
}

\date{Accepted 2025 June 11. Received 2025 May 14; in original form 2024 September 20; NORDITA-2024-027}

\pubyear{2024}

\begin{document}
\label{firstpage}
\pagerange{\pageref{firstpage}--\pageref{lastpage}}
\maketitle
\begin{abstract}
The low-frequency component of the upcoming Square Kilometre Array Observatory (SKA-Low) will be sensitive enough to construct 3D tomographic images of the 21-cm signal distribution during reionization. However, foreground contamination poses challenges for detecting this signal, and image recovery will heavily rely on effective mitigation methods. We introduce \serenet{}, a deep-learning framework designed to recover the 21-cm signal from SKA-Low's foreground-contaminated observations, enabling the detection of ionized (\hii{}) and neutral (\hi{}) regions during reionization. \serenet{} can recover the signal distribution with an average accuracy of 75 per cent at the early stages ($\overline{x}_\mathrm{HI}\simeq0.9$) and up to 90 per cent at the late stages of reionization ($\overline{x}_\mathrm{HI}\simeq0.1$). Conversely, \hi{} region detection starts at 92 per cent accuracy, decreasing to 73 per cent as reionization progresses. Beyond improving image recovery, \serenet{} provides cylindrical power spectra with an average accuracy exceeding 93 per cent throughout the reionization period. We tested \serenet{} on a 10-degree field-of-view simulation, consistently achieving better and more stable results when prior maps were provided. Notably, including prior information about \hii{} region locations improved 21-cm signal recovery by approximately 10 per cent. This capability was demonstrated by supplying \serenet{} with ionizing source distribution measurements, showing that high-redshift galaxy surveys of similar observation fields can optimize foreground mitigation and enhance 21-cm image construction.
\end{abstract}

\section{Introduction}

The 21-cm signal from the Epoch of Reionization (EoR) provides a unique window into the early Universe, revealing the processes of cosmic structure formation and the distribution of ionizing sources that transitioned the intergalactic medium (IGM) of our Universe from an initial cold and neutral to a final hot and ionized state \citep[e.g.][]{zaldarriaga200421,furlanetto2006cosmology,Pritchard201221cmCentury}. 
Several ongoing radio experiments, such as the Low-frequency Array \citep[LOFAR; e.g.][]{Mertens2020ImprovedLOFAR,Ghara2020ConstrainingObservations}, Murchison Widefield Array \citep[MWA; e.g.][]{tingay13,Trott2020DeepObservations} and Hydrogen Epoch of Reionization Array \citep[HERA; e.g.][]{HERA2022UpperPhaseIcontraints,HERA2023Improved} aim to observe this transition in our Universe.
The upcoming Square Kilometre Array\footnote{\url{https://www.skao.int/en}} Low-Frequency\footnote{\url{https://www.skao.int/en/explore/telescopes/ska-low}} component (SKA-Low) is expected to enhance our ability to observe this signal significantly, offering unprecedented sensitivity and spatial resolution \citep[e.g.][]{Koopmans2015TheArray}. However, extracting meaningful information from these observations presents substantial challenges, primarily due to the overwhelming foreground contamination and the inherent non-Gaussianity of the 21-cm signal. 

The foreground signal consists of galactic and extra-galactic components with various shapes, sizes and flux densities of the point and extended sources. At low radio frequencies, the foreground removal is a major obstacle, as the flux densities of this signal often far exceed the 21-cm signal by several orders of magnitude \citep[e.g.][]{ali2008foregrounds, Jelic2008Foreground, jelic2010realistic, ghosh2012characterizing}. Traditional foreground subtraction techniques struggle to manage the complex, diffuse nature of these foregrounds, leading to residuals that can obscure or distort the signal of interest. To address this, advanced methods are required to effectively isolate and recover the 21-cm signal from the observed data \citep[e.g.][]{Harker2009Nonparametric, Harker2010PowerCase, Liu2011AForegrounds, Liu2014EpochFormalism, Liu2014EpochReduction, chapman2012foreground, chapman2016effect, pober2016importance, Mertens2018Statistical21, hothi2021comparing, kern2021gaussian,acharya202421}.

In recent years, machine learning techniques have been consolidating their role as one of the most widely used tools employed by the cosmological and astrophysical scientific community for data analysis \citep[see e.g.][for on overview]{Dvorkin2022Machine,Lahav2023Deep}. In particular, U-Nets are becoming the standard in studies with tasks that require processing images or 3D data with structural correlations \citep[e.g.][]{He2015, Muenchmeyer2019, Perraudin2019, Gupta2020, Jeffrey2020, Guzman2021b, Gupta2021, Makinen2021, Bretonniere2021, Dong2021, Montefalcone2021, Han2021, Jeffrey2021, Casas2022, Li2022, Ostdiek2022a, Ostdiek2022b, Schaurecker2022, Piras2023}. In recent years, the field of 21-cm cosmology has employed  U-shaped networks (U-Nets) to tackle problems such as foreground contamination or performing other complex non-linear tasks \citep[e.g.]{Villanuevadomingo2020, GagnonHartman2021, Bianco2021segunet, Masipa2023, Hiegel2023, Bianco2023deep2, Kennedy2024, Shi2024, Sabti2024gen}. Although experiencing rapid success, the full potential of these networks still remains uncovered. Recently, \cite{Chen2024stability} investigated the strengths and limitations of this technique.

In our previous work, we focused on the detection of neutral regions with a segmentation network known as \segunet{} \citep[][hereafter Paper I]{} and tested this network's capability in the presence of foreground residuals \citep[][hereafter Paper II]{Bianco2023deep2}. In this paper, we want to take a step further and develop a network that also recovers the 21-cm signal. For this purpose, we go beyond the conventional development of U-Net and we propose a new architecture motivated by work from a completely different field. The authors of \cite{Tofighi2019prior} and \cite{Jurdi2020BBUnet} proposed a modified version of U-Net that includes an additional input to enhance their segmentation detection. We took inspiration from their work and decided to develop a new network with this alternative architecture to enhance the recovery of the 21-cm signal from foreground contamination images from SKA-Low tomographic images. Therefore, we constructed \recunet{} for recovering the signal. Apart from the standard elements of a U-Net \citep{Ronneberger2015} architecture (encode, decoder and skip layers), we construct an added component to the architecture that we named the \textit{Interception Convolutional Block}. A series of convolution and pooling layers processes an additional input that we refer to as the \textit{prior maps}, and the result intercepts the skipping layer of the U-Net to enforce prior information onto the network encoder and decoder.

Our new framework is the SEgmentation and REcover NEtwork (\serenet{}), developed for processing the SKA-Low multi-frequency tomographic observation of the 21-cm signal during the EoR. This framework comprises two neural networks and is combined with a preprocessing step, in our case, a PCA decomposition, that helps remove the galactic foreground largest contribution, which results in an image with some residual foreground contamination and systematic noise. This step provides the starting point for \serenet{} framework. Its first network, \segunet{} \citepalias{Bianco2021segunet}, processes the residual image into a binary map, which indicates the neutral (\hi{}) and ionized (\hii{}) regions. This binary map is employed as additional input in \recunet{} by the \textit{interception convolution block} along with the residual image to recover the underlying 21-cm signal distribution in the tomographic data.

We perform a complete study assessing the new architecture and the advantage of including different binary prior maps in recovering 21-cm signals from foreground-contaminated EoR data. We demonstrate that by using a binary field derived from the volume-averaged neutral fraction field $x_\mathrm{HI}(\mathbf{r}, z)$, the \serenet{} frameworks achieve more stable predictions and a better 21-cm signal recovery. Furthermore, we demonstrate that any meaningful prior information could be used without requiring retraining the network. Finally, we show that \serenet{} could potentially be employed for synergy observations between SKA-Low and high redshift galaxy observations, such as the James Webb Space Telescope (JWST), Euclid and Nancy Grace Roman Space Telescope. This synergy would be potentially beneficial to both experiments by enhancing 21-cm recovery and understanding the relation between the early sources of reionization and properties of the IGM \citep[e.g.][]{Zackrisson2020Bubble6}.

This paper is structured as follows: \S\ref{sec:21cm_simulation} provides an overview of our simulated 21-cm signal dataset. In \S\ref{sec:ML}, we introduce our neural network framework, \serenet{}, designed to mitigate foreground contamination. \S\ref{sec:results} presents our results, and we conclude in \S\ref{sec:conclusion}. In Appendix~\ref{ap:application}, we describe our method for applying \serenet{} to larger images than those used for training. In this work, we assume a flat $\Lambda$ cold dark matter cosmology, with the following cosmological parameters: $\Omega_b = 0.046$, $\Omega_m = 0.27$, $H_0 = 100h = 70~{\rm km s^{-1} Mpc^{-1}}$, $\sigma_8 = 0.82$, and $n_s = 0.96$ \citep[e.g.][]{Komatsu2011Seven-yearInterpretation, PlanckCollaboration2018}.

\section{21-cm Simulation Dataset}\label{sec:21cm_simulation}
Radio telescopes observe the 21-cm signal as differential brightness temperature, which is given as \citep[e.g.][]{Pritchard201221cmCentury}
\begin{align}\label{eq:dTb}
	\delta T_\mathrm{b} (\mathbf{r},z) = 27 ~\mathrm{mK} \left(\frac{\Omega_b h^2}{0.022}\right) &\left(\frac{0.144}{\Omega_m h^2} \frac{1+z}{10}\right)^{\frac{1}{2}} \nonumber \\
	&x_\mathrm{HI}(\mathbf{r},z)(1+\delta(\mathbf{r},z)) \left(1-\frac{T_\mathrm{CMB}(z)}{T_\mathrm{S}(\mathbf{r},z)}\right) \ ,
\end{align}
where $x_\mathrm{HI}$, $\delta$, $T_\mathrm{S}$, and $T_\mathrm{CMB}$ are the neutral fraction, matter over-density, spin temperature, and cosmic microwave background (CMB) temperature, respectively. The spin temperature can be assumed to be saturated ($T_\mathrm{S}\gg T_\mathrm{CMB}$) at $z\lesssim 9$ \citep[e.g.][]{Ross2017SimulatingDawn, Ross2021CosmicDawn, HERA2022UpperPhaseIcontraints}. However, in this study, we explore up to $z\approx 11$. We defer the proper inclusion of realistic spin temperature fluctuations to the future as this calculation requires more computing resources. Our framework is still reliable at $z\lesssim 9$ as the trained network is independent of redshift. 

This section describes how we created the dataset to train our new foreground mitigation framework. \S\ref{sec:21cmFast} focuses on the reionization simulations, and \S\ref{sec:obs_effect} presents the modelling of the telescope effects and foreground contamination.

\begin{figure*}
	\includegraphics[width=0.9\textwidth]{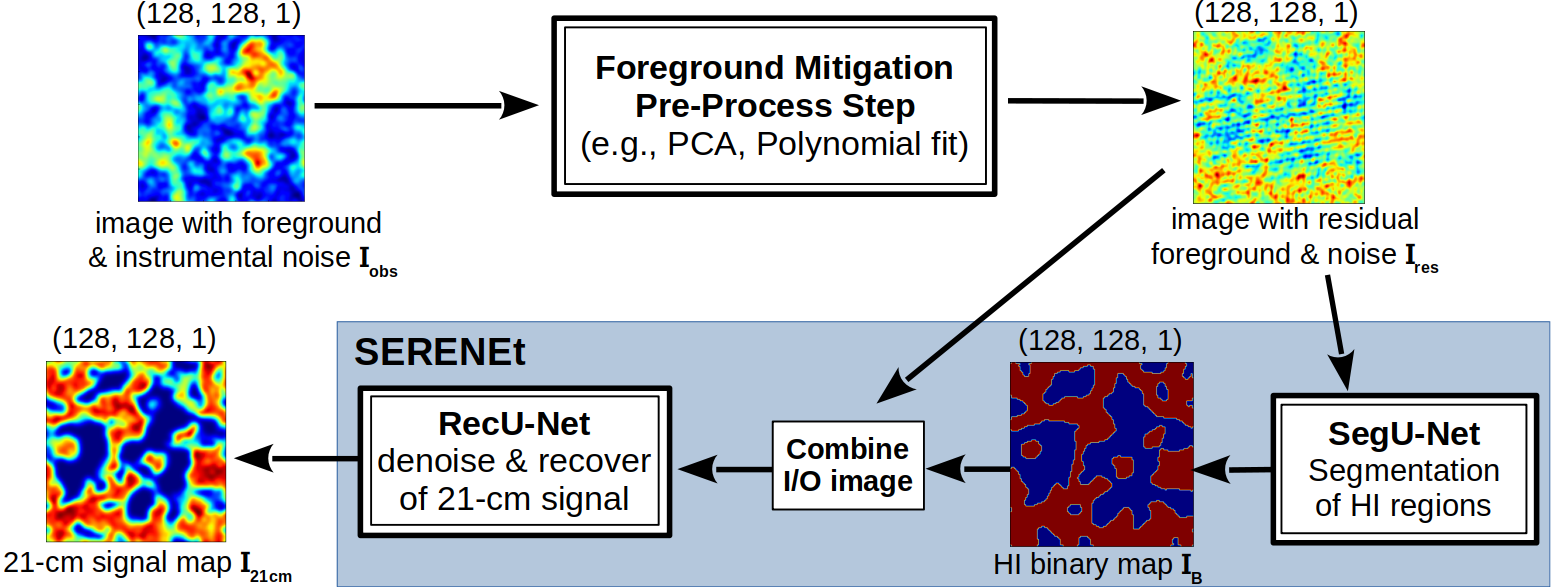}
	\caption{Schematic diagram of the \serenet{} pipeline. The first step consists of pre-processing the observed image, $I_\mathrm{obs}$, to remove the largest contribution of the foreground contamination, resulting in a residual image, $I_\mathrm{res}$. The second step involves identifying neutral regions with the \segunet{} architecture. The segmentation provides a binary map, $I_\mathrm{B}$, that, combined with the residual image, provides the input of the third step, which consists of recovering the 21-cm signal with the \recunet{} architecture. The combination of the two networks constitutes \serenet{} that constructs the final image.}
	\label{fig:serenet_pipeline}
\end{figure*}

\subsection{Modelling structure formation and reionization }\label{sec:21cmFast}
We employ a semi-numerical code, named \textsc{21cmFAST} \citep[e.g.][]{Mesinger2011,Murray202021cmFAST} that allows simulating both the cosmic structure formation and ionization of the IGM. The large-scale dark matter distribution at any cosmic time is determined using second-order Lagrangian Perturbation theory \citep[2LPT;][]{Scoccimarro1998NonlinearPerturbations} in simulation volumes on grids of $128^3$ and length of 256 Mpc along each direction. While several more accurate modelling frameworks, such as \texttt{pyC$^2$Ray} \citep{hirling2023pyc} and \texttt{CRASH} \citep{Maselli2003CRASH:Scheme}, exist, we chose this semi-numerical code as it is computationally inexpensive enough to allow us to create a large dataset for training our deep learning model \serenet{}.

The reionization process is modelled using the excursion-set-based formalism initially proposed by \citet{Furlanetto2004} and put on the simulation grid in \citet{Mesinger2007EfficientReionization}. In this formalism, a cell is ionized if the following condition is satisfied,
\begin{eqnarray}
	\zeta f_\mathrm{coll} > 1 ,
\end{eqnarray}
where $\zeta$ is the ionizing source efficiency and $f_\mathrm{coll}$ is the collapse fraction of dark matter haloes of mass more than $M_\mathrm{min}$. This quantity is estimated at this cell assuming a spherical top-hat filter of radius $R_\mathrm{s}$. Several $R_\mathrm{s}$ values are tested until a maximum value of $R_\mathrm{mfp}$. 

We created a dataset of 10,000 astrophysical models by varying the three astrophysical parameters $(\zeta, M_\mathrm{min}, R_\mathrm{mfp})$ for training the networks. This dataset was used in \citetalias{Bianco2021segunet} and \citetalias{Bianco2023deep2}. These models were produced at redshifts between 11 and 7 with a redshift width of 0.2. For more details about the astrophysical models, see \citetalias{Bianco2021segunet}. We then simulate an additional 1,500 simulations for the validation test employed during the training of \serenet{}. Finally, we simulate 300 more as the testing sets discussed in \S\ref{sec:ML} and refer to it as the \textit{random dataset}. From the testing set, we select one realization that we employ as a \textit{fiducial} model. We should note that the neural network trained to learn the patterns in the 21-cm signal images is not dependent on the exact values of the astrophysical parameters \citepalias{Bianco2021segunet, Bianco2023deep2}. In \S\ref{ap:datasize}, we investigate the implications of the dataset size on the accuracy and performance of our model.

\subsection{Modelling the observational effects}\label{sec:obs_effect}
We constructed synthetic observations of SKA-Low by simulating the \textit{uv} tracks using the planned antenna locations, conforming to the SKAO configuration coordinates plan \citep{skao2016config}. The observations assume a total observation time of 1000 hours, with data collected for 6 hours each day. This extended observation time helps to reduce instrumental noise sufficiently to resolve meaningful features in 21-cm signal images \citep{Giri2018OptimalObservations}. Additionally, we consider the antenna distribution contained within a maximum diameter of 2 km around the core of SKA-Low that reduced the resolution of the 21-cm signal image data \citep{Mellema2015HISKA, Giri2018BubbleTomography}.

We modelled the foreground contamination in the 21-cm signal data following the procedure outlined in \citet{Choudhuri2014foreground}. We refer to section 2.3 of \citetalias{Bianco2023deep2} for a summary of this procedure. This work focuses solely on the Galactic synchrotron emission, assuming that point sources have been perfectly removed. Although this emission is several orders of magnitude stronger than the 21-cm signal, it exhibits a smooth frequency dependence. Several foreground mitigation methods exploit this characteristic \citep[e.g.,][]{Harker2009Nonparametric, chapman2016effect, Mertens2018Statistical21}. All methods used to produce synthetic observations in this study are publicly available in \texttt{tools21cm} \citep{Giri2020t2c}.

\section{Machine Learning Application} \label{sec:ML}
This section presents our pipeline, which includes the deep-learning approach we employ to mitigate foreground contamination. In \S\ref{sec:pipeline}, we explain the overall pipeline that combines any classical foreground mitigation method with two convolutional U-shaped networks (U-Nets). In \S\ref{sec:network_architecture}, we discuss how we employ this novel neural network architecture that can input more than one image data. Finally, \S\ref{sec:metric_uncertaint} presents the metrics for quantifying the accuracy of the framework.

\begin{figure*}
	\includegraphics[width=0.95\textwidth]{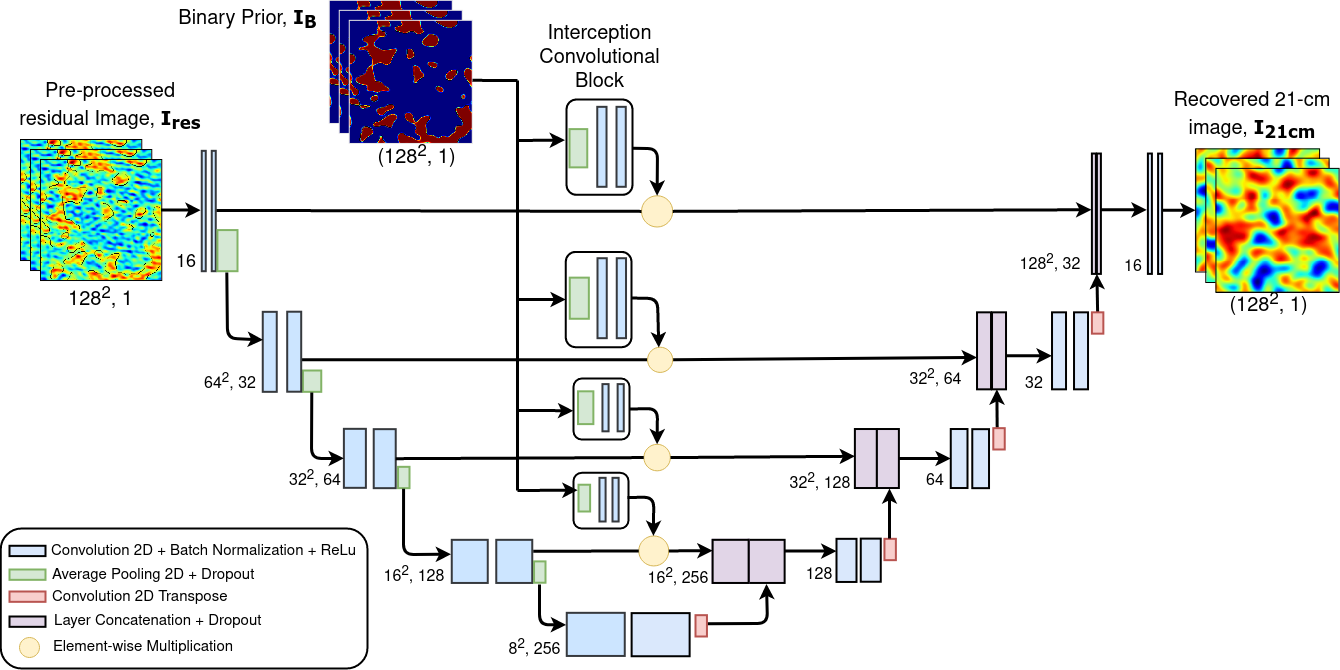}\vskip-2mm
	\caption{Architecture of the RecU-Net model. The standard U-shape is modified by introducing \textit{Interception Convolutional Blocks}, which consist of a series of pooling and convolutional operations that intercept and combine with the output of the skip connections. In this case, we use binary images from \segunet{} predictions, but any data can be used as prior in this architecture.}
	\label{fig:network_architecture}
\end{figure*}

\subsection{SERENEt pipeline}\label{sec:pipeline}
Several previous works have demonstrated that foreground mitigation can be improved by combining machine learning techniques with standard mitigation approaches such as principal component analysis (PCA) and polynomial fitting \citep[e.g.][]{Li2019Sep, Makinen2021deep}. Therefore, our pipeline follows this approach and is divided into three main steps. In \autoref{fig:serenet_pipeline}, we show a schematic diagram of this pipeline. In the first step of our pipeline, we pre-process the foreground-contaminated image, $I_{\rm obs}$, of mesh-size $128^2$, with a simple PCA method\footnote{We employ the module implemented in the \texttt{SciPy} software package that can be found at \url{https://docs.scipy.org}.}. This method models the foreground by applying PCA at each pixel along the frequency direction. As the foreground signal is expected to be smooth along frequency, the components with the most significant contributions should contain the foreground signal. The choice of the number of PCA modes corresponding to the foreground signal affects the level of residuals \citep[e.g.][]{Harker2009Nonparametric}. This study removes the first four modes from a PCA decomposition in the tomographic dataset and reconstructs the image, $I_{\rm res}$, with the remaining components. This step allows us to remove the significant contribution of the foreground signal. However, as discussed next, the neural network will handle systematic noise and the remaining residual foreground signal.

The residual image is provided as the input to the \serenet{} that consists of two U-Nets, namely \segunet{} and \recunet{}. The former is a segmentation network already presented in our previous works \citepalias{Bianco2021segunet, Bianco2023deep2}. This network takes the residual image $I_{\rm res}$ and provides a binary map, $I_{\rm B}$, that identifies a region with neutral hydrogen (with value 1) or ionized (with value 0).  We should note that \segunet{} was trained to assign cells that are more than $50$ per cent ionization fraction to be ionized. The second network, \recunet{}, uses a modified version of the U-Net architecture \citep{Ronneberger2015} that can take an additional input containing prior information. In our case, we use the output of \segunet{}, which are the {\sc Hii} binary maps, as a second input that will function as a prior map, indicating regions in the sky map that are expected to emit 21-cm signal due to the presence of neutral hydrogen in the IGM. The two inputs are combined by \recunet{} to recover the simulated 21-cm signal image $I_{\rm 21cm}$.

We will explain the architecture of the second network in more detail in the next subsection (\S\ref{sec:network_architecture}). In this work, we limited ourselves to using the binary prediction of our segmentation network as prior information. However, we will show in \S\ref{sec:galaxy} the possibility of combining SKA-Low tomographic images with other high-redshift experiments (e.g., James Webb Space Telescope, Euclid, Nancy Grace Roman Space Telescope, and Extremely Large Telescope) to enhance the recovery process of the 21-cm signal when these observations manage to cover a good portion of SKA-Low field of view.

\subsection{Neural Network Architecture}\label{sec:network_architecture}
The current deep learning-based framework inherits \segunet{} architecture from our previous work \citepalias{Bianco2021segunet, Bianco2023deep2} and further develops a methodology to recover 21-cm signal images from data contaminated with foreground signals. We propose an additional network architecture called \recunet{}, illustrated in \autoref{fig:network_architecture}. This modified version of the classical and widely used U-Net architecture includes a second input image to enhance the reconstruction of the 21-cm signal from the residual image, $I_{\rm res}$.

In \recunet{}, the first input, $I_{\rm res}$, is fed into the standard input layer of the U-Net architecture. The second input, consisting of binary maps from \segunet{}, undergoes a pooling operation that resizes the image to the corresponding level of the U-Net, followed by two convolutional layers: \texttt{Pool+2*(Conv2D+BN+Activ)}. Similar to our previous work, we employ \texttt{ReLU} activation function \citep{Agarap2019ReLU}. We refer to this part of the network architecture as the \textit{Interception Convolutional Block} because the output of these layers intercepts and combines with the skip-connection of the U-shaped network through element-wise multiplication.

This new approach was inspired by the work of \cite{Tofighi2019prior} and \cite{Jurdi2020BBUnet}. We specifically based our model on the latter, which employed a similar architecture in a different domain to improve segmentation prediction. In our case, we adapted their model by replacing the target shape and position estimates with the binary predictions of our segmentation network, as detailed in \S\ref{sec:pipeline}. The intention is to use the segmentation map as prior information indicating where we expect the emission of 21-cm signals in the sky image. Therefore, we refer to the prediction of \segunet{} as the \textit{21-cm image binary prior} or simply \textit{prior}.
\begin{figure*}
	\includegraphics[width=\textwidth]{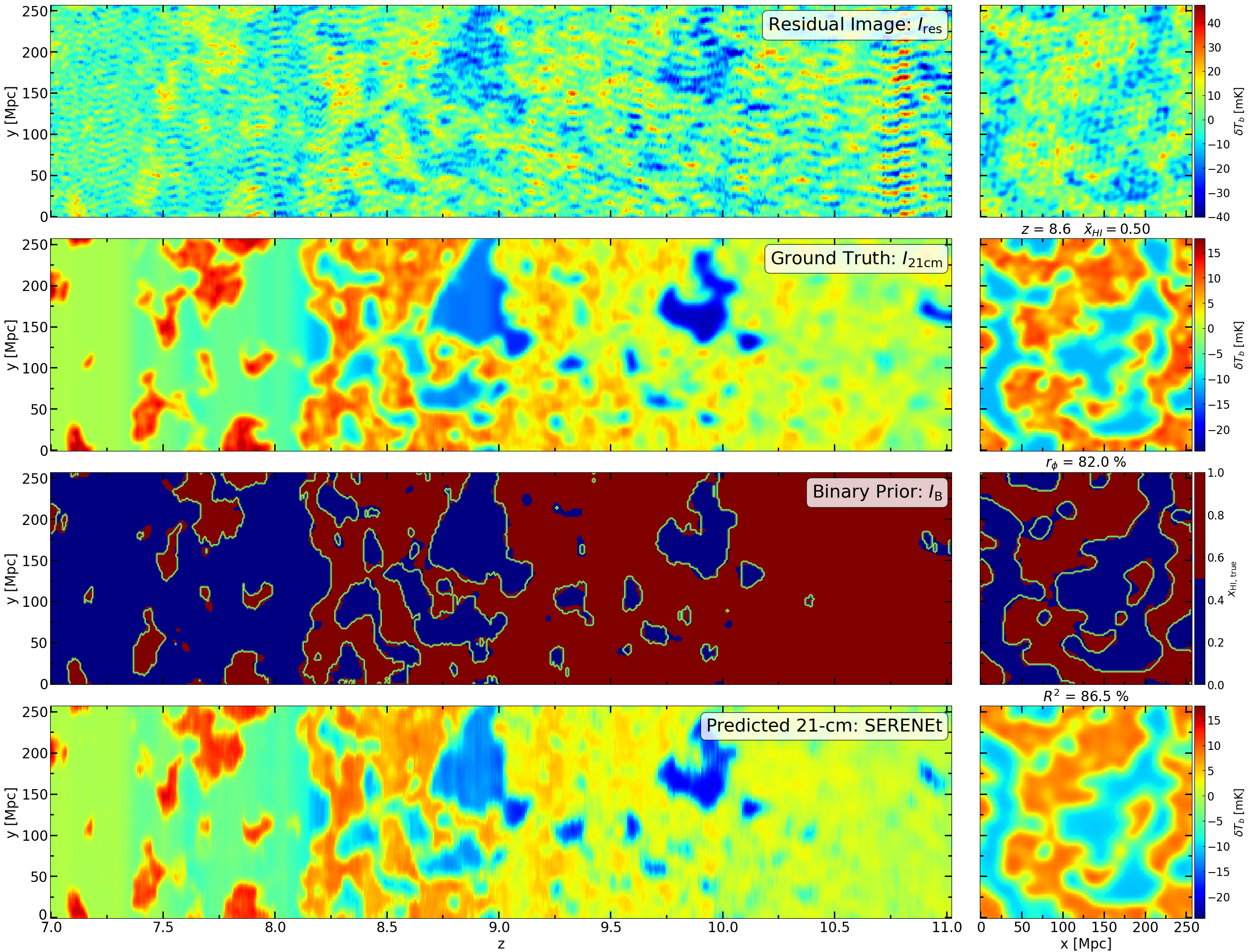}\vskip-2mm
	\caption{Example of different datasets from the \textit{fiducial} model. The left column shows the redshift evolution of the field along the y-axis, while the right column displays the 21-cm fluctuation when the IGM is 50 per cent ionised. The top panels show the residual input, $I_\mathrm{res}$. The second row shows the simulated 21-cm signal, $I_\mathrm{21cm}$, which is the target for \serenet{}. The third row presents the binary map, $I_\mathrm{B}$, generated by applying a threshold to identify regions as either neutral or ionised, with contours indicating the \segunet{} prediction. We show the \serenet{} prediction in the last row, using its results as prior information. \serenet{} predicts the signal with an $R^2$ score of 86.5 per cent.}
	\label{fig:visual_fiducial}
\end{figure*}

\subsection{Metrics for Training and Validation}\label{sec:metric_uncertaint}
In this paper, we employ the coefficient of determination, $R^2$, in the regression performed with \serenet{} when quantifying the different results of recovering the 21-cm images, $\hat{y}$ and the ground truth, $y$. This metric is defined as \citep[e.g.][]{draper1998applied, montgomery2021introduction}:
\begin{equation}\label{eq:r2score}
    R^2 (y, \hat{y}) = 1 - \frac{\sum_i (y_i - \hat{y}_i)^2}{\sum_i (y_i - \bar{y})^2}\,,
\end{equation}
where $\bar{y} = \frac{1}{N}\sum_i y_i$ is the average of the image at a given redshift, and the sum is over all the pixels. In our case, this quantity corresponds to the volume-averaged ionized fraction, $\overline{x}_\mathrm{HII}$, at each redshift or observed frequency. The numerator is the mean squared error of the network prediction, and it is divided by the residual sum of squares concerning the averaged value in the dataset. The $R^2$ metric has a maximum value of $1$, which indicates a positive correlation. Values near zero show that the prediction is completely random and thus uncorrelated to the target, while values below zero indicate the worst result. We prefer to use the $R^2$ score for both validation and during the training, as this metric is more intuitive than the regular mean square error or mean absolute error, as it can be expressed in percentage. Moreover, it is more informative in regression analysis compared to the latter two metrics \citep{Chicco2021}.

Another important metric we presented in our previous works \citepalias{Bianco2021segunet, Bianco2023deep2} is the Matthews correlation coefficient (MCC), $r_\phi$. This metric is employed in image classification and segmentation, and, in our case, we use it to evaluate \segunet{} binary predicted maps. We define this metric as \citep[e.g.][]{bishop2006pattern},
\begin{equation}\label{eq:mmc}
    r_{\phi} (y, \hat{y}) = \frac{ TP \cdot TN - FP \cdot FN }{ \sqrt{ (TP+FP) (TP+FN) (TN+FP) (TN+FN)}} \ ,
\end{equation}
where $TP = \sum_i y_i \hat{y}_i$ indicates the true positive, $TN = \sum_i (1-y_i)(1-\hat{y}_i)$ is the true negative, $FP = \sum_i y_i (1-\hat{y}_i)$ is the false positive, and $FN = \sum_i (1-y_i) \hat{y}_i$ is the false negative. $1$ indicates a neutral region in our binary maps, while $0$ is ionized. Therefore, in our case, $TP$ ($TN$) indicates the number of pixels in the image that are correctly identified as neutral (ionized). Instead, $FP$ and $FN$ quantifies the corresponding wrong classification. This metric ranges between $-1\leq r_{\phi}\leq1$; a positive value indicates a correlation between the predicted binary map and the ground truth image. A negative value instead indicates anti-correlation, while values close to zeros indicate a completely random classification.
\begin{figure*}
	\includegraphics[width=\textwidth]{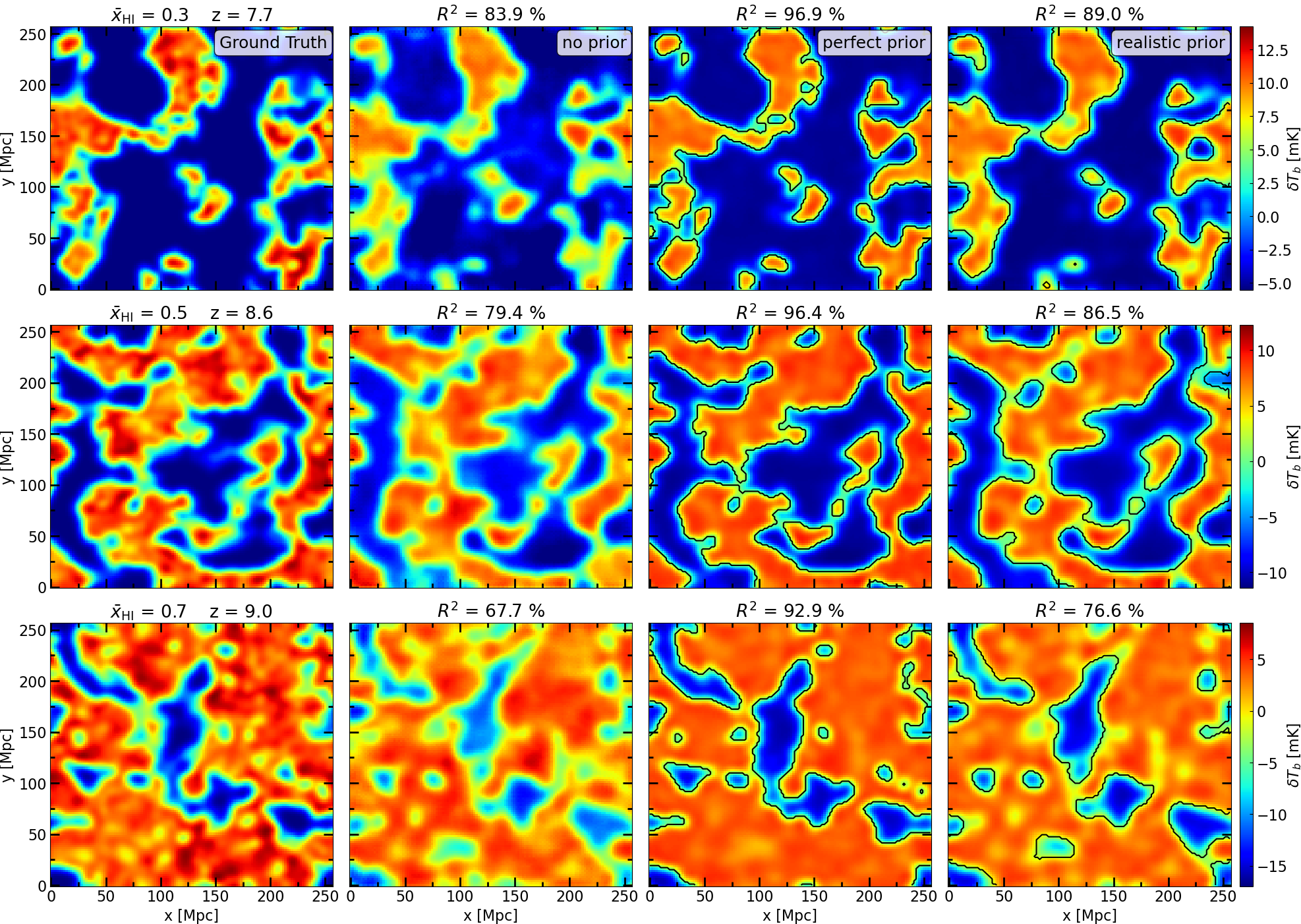}\vskip-2mm
	\caption{Visual comparison of the predicted 21-cm signal for different setups (columns) at various redshifts (rows). From top to bottom, the rows correspond to $z = 7.689$, $8.631$, and $9.0$, representing reionization milestones $\overline{x}_\mathrm{HI,,true} = 0.7$, $0.5$, and $0.3$, respectively. From left to right, the columns show the ground truth and predictions from \serenet{} for the \textit{no prior}, \textit{perfect prior}, and \textit{realistic prior} cases. In the last two cases, we include contours of the binary prior, corresponding to the ground truth binary maps and \segunet{} predictions. While \serenet{} can recover the 21-cm signal images without prior information, the accuracy improves significantly by including priors. Although the \textit{perfect prior} case yields the best results, we can observe a substantial improvement with the \textit{realistic prior} at all redshifts.}
	\label{fig:visual_comparison}
\end{figure*}
\section{Results} \label{sec:results}
In this section, we analyse the prediction of our networks. In particular, we consider three cases. In the first case, we consider 21-cm images reconstructed by a simple U-Net. This is equivalent to employing \recunet{}, but without the \textit{Interception Convolutional Block}; in this case, we want to test if the modified architecture of \recunet{} provides an actual advantage. We refer to this setup as the \textit{no prior} test. We consider the modified architecture in the other two cases, as illustrated in \S\ref{sec:network_architecture}. The former considers the perfect scenario where the binary prediction, thus the prior, from \segunet{} obtains a perfect score. We refer to this setup as the \textit{perfect prior} test, and in this case, we employ the actual ground truth binary maps in the training and validation of \recunet{}. Our last test is the \textit{realistic prior} setup. Here, we employ the trained \segunet{} as presented in \citetalias[][]{Bianco2023deep2} and provide a realistic scenario where the recovery of the 21-cm image by \recunet{} will also depend on the accuracy of the binary prior predicted by \segunet{}.

First, we will show the ability of \serenet{} to recover the 21-cm signal image data in \S\ref{sec:visual_eval}. Next, in \S\ref{sec:eff_prior}, we study the improvement gained by providing any prior information. We statistically evaluate the accuracy of \serenet{} in \S\ref{sec:stat_eval} and the study of the recovery of the 21-cm power spectra in \S\ref{sec:recover_ps}. In \S\ref{sec:eos_dataset}, we employ our framework on the 21-cm signal measured for a larger field of view compared to our training set. Finally, we test the ability of \serenet{} to include information in galaxy surveys in \S\ref{sec:galaxy}.

\subsection{Recovered 21-cm Signal Spatial Domain}\label{sec:visual_eval}
As mentioned earlier, one of the main advantages of the SKA-Low experiment is its ability to produce tomographic images of the 21-cm signal \citep[e.g.][]{Mellema2015HISKA, Ghara2017ImagingSKA, Giri2018BubbleTomography, Giri2018OptimalObservations}. These images will allow us to extract information beyond the Gaussian statistics contained in this signal \citep[e.g.][]{Giri2019NeutralTomography, Giri2019Position-dependentReionization, Watkinson2019TheHeating}. Therefore, we will assess the accuracy of the 21-cm images recovered by \serenet{} from the contaminated SKA-Low data.

First, in \autoref{fig:visual_fiducial}, we visually evaluate the recovery of our \textit{fiducial} model. The right columns display maps corresponding to the time when half of the hydrogen in the IGM is ionised, i.e., $\overline{x}_\mathrm{HI} = 0.5$. In the \textit{fiducial} model, this occurs at $z=8.6$. The left columns show the redshift evolution of the 21-cm signal along a specific declination. We show the residual input $I_{\rm res}$ on the top row that our PCA foreground mitigation method provides. The second row shows the differential brightness temperature calculated with \autoref{eq:dTb} and created as explained in \S\ref{sec:21cm_simulation}. This field is the ground truth of \serenet{} and the target prediction of \recunet{}, $I_\mathrm{21cm}$. Comparing these two rows, we observe that the foreground contamination and noise residuals are several times larger than the expected 21-cm signal. Although large-scale features in the 21-cm signal distribution can be identified in the top row, some artificial features remain in the redshift direction due to the pre-processing.

In the third row, we display binary maps for ionised and neutral regions, $I_\mathrm{B}$, which were produced by applying a threshold of $x_\mathrm{HI}=0.5$ to the ionisation map, smoothed to the resolution of SKA-Low with a maximum baseline of 2 km. This field is the target of \segunet{}, and green contours indicate the corresponding prediction by \segunet{}. It is important to note that \segunet{} is applied on the signal containing residual foreground (top row). This prediction is later used as prior information in \serenet{} for the \textit{realistic prior}. Additionally, we will consider \textit{perfect prior} scenario where the binary maps are used directly. The fourth row shows the \serenet{} prediction for the \textit{realistic prior} scenario. When compared to the second panel, it is evident that all the features are reconstructed accurately. In the following subsections, we will discuss further how the binary maps, used as prior information as illustrated in \autoref{fig:network_architecture}, impact the recovery of 21-cm images in \serenet{}.

\subsection{Effect of the Prior on the 21-cm Signal Predictions}\label{sec:eff_prior}
The primary motivation for implementing the new architecture, as illustrated in \S\ref{sec:network_architecture}, is to improve the recovery of the 21-cm signal in foreground-contaminated observations by incorporating prior information about the ionised regions. In this section, we evaluate the effectiveness of this prior to the \textit{fiducial} model.

In \autoref{fig:visual_comparison}, we show the recovered 21-cm signal at different stages of reionization at redshifts $z = 9.0$, $8.6$, and $7.7$ (different rows), corresponding to $\overline{x}_\mathrm{HI, true} = 0.7$, $0.5$, and $0.3$. In this figure, the first column represents the ground truth, while the remaining columns display three different test cases.
The second column shows the most basic scenario, \textit{no prior}, which uses a standard U-shaped architecture \citep{Chen2024stability}, as illustrated in \autoref{fig:network_architecture}, but without the \textit{Interception Convolutional Block}. The third column presents the \textit{perfect prior} case, where the modified architecture described in \S\ref{sec:network_architecture} is used with the ground truth binary maps as prior. This case assumes that \segunet{} perfectly identifies the neutral regions, which is challenging to achieve, representing the best possible result within the \serenet{} framework. The fourth column shows the \textit{realistic prior} case, where we use the predicted binary maps from \segunet{} as priors. Black contours in the last two columns indicate the corresponding prior maps.

The first observation is that while the simple U-shaped architecture without binary prior maps recovers the 21-cm signal with satisfactory accuracy ($0.65 < R^2 < 0.85$), it underperforms compared to cases that utilize the prior $I_\mathrm{B}$. As expected, the \textit{perfect prior} case yields the best predictions, with $R^2 > 0.9$ for the three reionization stages presented in \autoref{fig:visual_comparison}. In contrast, the \textit{realistic prior} provides results that fall between the other two cases.

\begin{figure}
	\includegraphics[width=0.98\columnwidth]{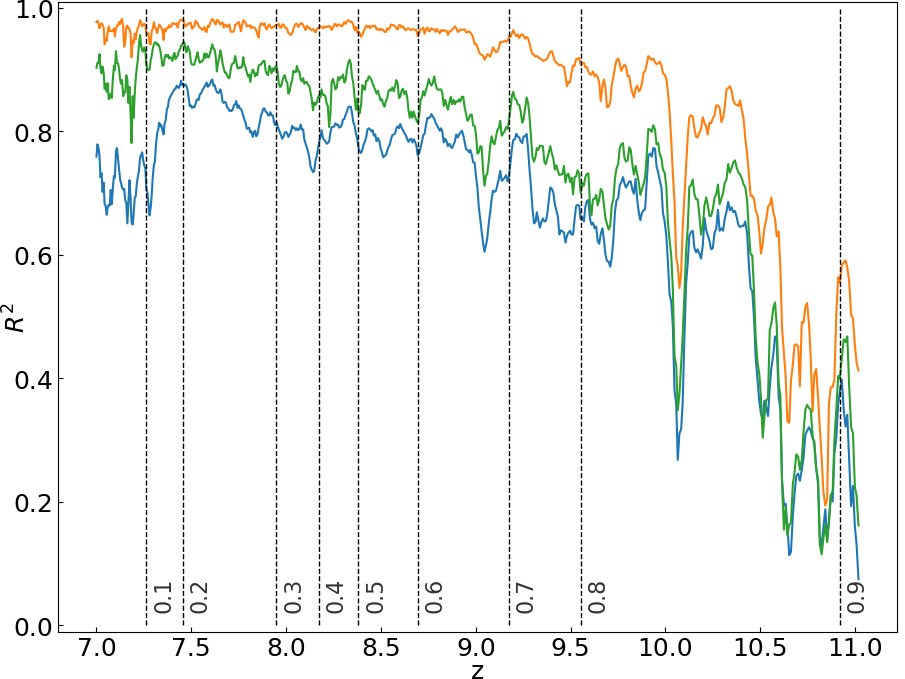}\vskip-2mm
	\caption{Redshift evolution of the $R^2$ score for the prediction by \segunet{} in the \textit{fiducial} model. The result shows different scenarios: \textit{no prior} (blue line), \textit{perfect prior} (orange line), and \textit{realistic prior} (green line).}
	\label{fig:stat_fiducial}
\end{figure}

\begin{figure*}
	\includegraphics[width=\textwidth]{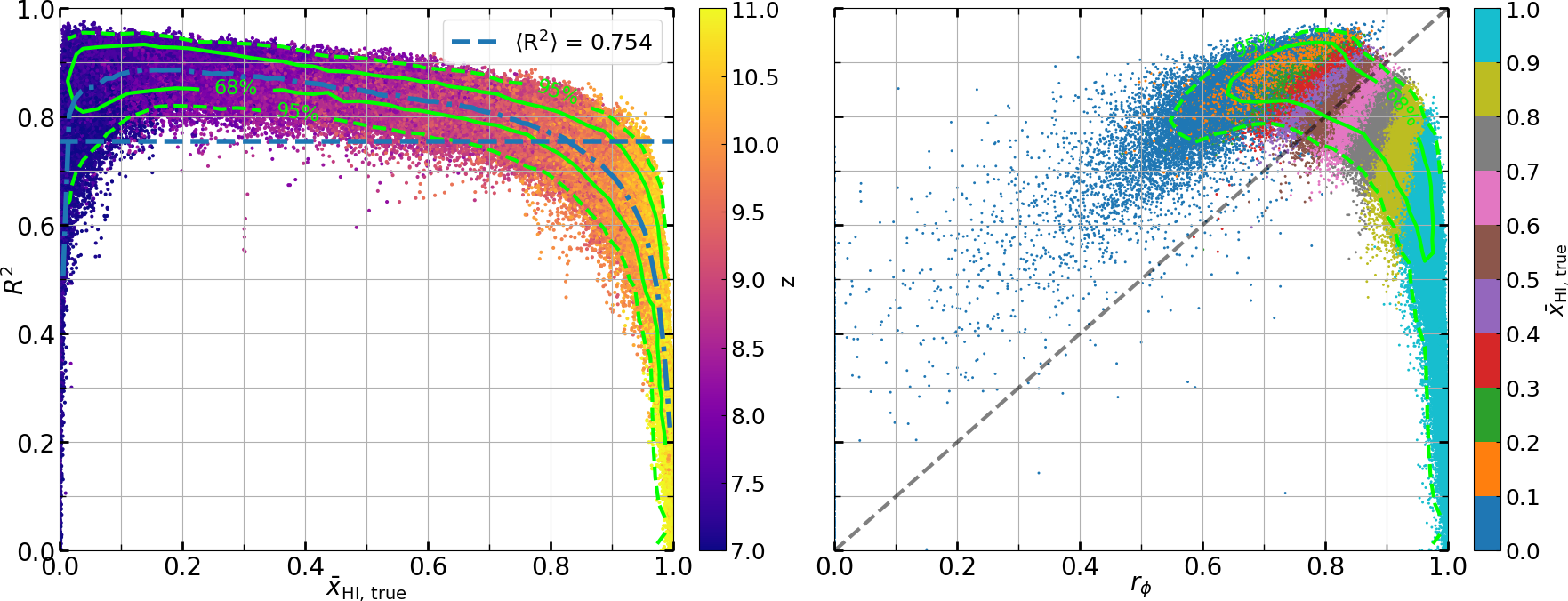}\vskip-2mm
	\caption{Statistical results of \serenet{} predictions for the \textit{realistic prior} case on the \textit{random dataset}. \textit{Left panel}: correlation between the $R^2$ score and the true volume-averaged neutral fraction $\overline{x}_{\mathrm{HI,true}}$. Each point represents 21-cm signal data at the corresponding redshift, indicated by the colour bar. \textit{Right panel}: correlation between the $R^2$ score and the binary score $r_\phi$ of the realistic prior predicted by \segunet{}. The colour indicates the $\overline{x}_{\mathrm{HI,true}}$ of the image. Ideally, all points should be close to the top right corner, representing a perfect full pipeline. However, we observe that data with a high $\overline{x}_{\mathrm{HI,true}}$ has good \segunet{} predictions but less accurate \serenet{} results, while data with a low $\overline{x}_{\mathrm{HI,true}}$ shows poor \segunet{} predictions but accurate \serenet{} results. Both panels show 68 and 95 per cent confidence intervals with green solid and dashed contours.}
	\label{fig:stat_dataset}
\end{figure*}

Moreover, it is noticeable that in the images recovered for the \textit{perfect prior} case, there are almost no fluctuations within the regions emitting 21-cm signals. This feature is particularly evident at higher redshifts. For example, the variations in the signal around the regions $(x, y) \approx (200, 200)\,\rm Mpc$ or $(x, y) \approx (125, 25), \rm Mpc$ in the ground truth at $z = 9.0$ are absent in the \textit{perfect prior} case. This suppression of fluctuations seems to be due to the reliance on the perfect prior, which mainly influences the regression process, thereby reducing signal variability.

A similar pattern is observed in the \textit{realistic prior} case, although the fluctuation suppression is less pronounced. An imperfect binary prior can sometimes benefit the regression step with \recunet{}, as illustrated by the advantage provided when \segunet{} mistakenly detects a bubble at $(x, y) = (100, 40)\, \rm Mpc$ at $z = 9.0$. This erroneous detection helps recover a depression in the signal that is not captured when using the binary ground truth as prior. However, errors in the binary predictions can also have detrimental effects. For instance, the realistic prediction mistakenly interprets the large percolated region at $(x, y) = (120, 90)\, \rm Mpc$ at $z = 9.0$ in the ground truth as two separate bubbles.

The advantage of the prior is particularly evident from the redshift evolution of the $R^2$ score, shown in \autoref{fig:stat_fiducial}. The $R^2$ score for the \textit{no prior} case (blue line) is substantially lower than that for the other tests—approximately $5-35$ per cent below the \textit{perfect prior} (orange line) and up to $25$ per cent below the \textit{realistic prior} (green line). Furthermore, the accuracy of the \textit{perfect prior} scenario remains constant at $R^2 \approx 0.97$ until $z \simeq 9$, when $\overline{x}_\mathrm{HI} \gtrsim 0.7$. In contrast, the accuracy in the other cases gradually decreases at higher redshifts. This trend depends on the increasing noise level, but mainly on the number of components we remove in the PCA pre-processing. Removing more components allows for a better mitigation of the foreground at higher redshift, lightening the accuracy decrease at higher redshift but increasing the portion of the 21-cm signal removed at lower redshifts. Following the approach in \citetalias{Bianco2023deep2}, we removed the first four components, which contained the most significant contribution of the foreground contamination, and as it provides a balanced accuracy between high and low redshift.

Moreover, in \autoref{fig:stat_fiducial}, one can notice a sudden drop in accuracy at different redshifts in all the scenarios, in the \textit{fiducial} model, the most evident one occurs at $z\sim10$. These events are due to numerical artifacts in the imperfect PCA foreground decomposition, as we can observe them in the \textit{perfect prior} only at high redshift, $z>9$. We statistically explore these trends further in the next section, \S\ref{sec:stat_eval}.

\subsection{Statistical Evaluation of \serenet{}}\label{sec:stat_eval}
In this section, we statistically quantify the ability of \serenet{} to recover the 21-cm signal from the residual images, $I_{\rm res}$, using a large testing dataset. We focus primarily on the \textit{random dataset}, a testing set of 300 data points, and the \textit{realistic prior} case, as these represent the most likely scenarios for actual observations.

In the left panel of \autoref{fig:stat_dataset}, we show the $R^2$ score (\autoref{eq:r2score}) against the true neutral fraction, $\overline{x}_{\mathrm{HI,true}}$. Each point represents a 21-cm image at a given redshift, as the colorbar indicates. As expected, the images at the lowest ($z \sim 7$) and highest ($z \sim 11$) redshifts show a decline in the $R^2$ score due to the inefficiency of PCA method modelling the foreground at the boundaries of the lightcone data that we consider. The average accuracy on the entire \textit{random dataset} is approximately $\left<R^2\right> = 0.754$. Meanwhile, the blue dash-dotted line indicates the mean score as a function of volume-averaged neutral fraction, $R^2(x_\mathrm{HI})$. In \autoref{tab:accuracy}, we show the accuracy of the recovered 21-cm maps and the corresponding predicted binary prior to the different reionization milestones; in bold, we highlight the bin corresponding to the visual example of the \textit{fiducial} model. The contour lines show the $68$ per cent (green solid line) and $95$ per cent (green dashed line) confidence intervals of the data, with a corresponding average accuracy of $\left<R^2\right> = 0.763$ and $\left<R^2\right> = 0.760$, respectively.

In the right panel of \autoref{fig:stat_dataset}, we show the correlation plot between \recunet{}'s accuracy in recovering the 21-cm signal ($R^2$) and its dependency on the quality of the binary prior map predicted by \segunet{} ($r_\phi$), which is defined by \autoref{eq:mmc}. The colour of the data points indicates the ground truth neutral fraction at intervals of $0.1$. A dashed black line represents the expected behaviour of the $R^2 - r_\phi$ relation, as we expect that the quality of the binary prior will affect the prediction of the 21-cm signal. Ideally, most points would be located at the top right of the plot. In our case, $68$ per cent of the recovered 21-cm images have $r_\phi \geq 0.65$ and $0.52 \leq R^2 \leq 0.92$, indicating that most of the 21-cm signals from different stages of the reionization history are accurately recovered. The $95$ per cent contour includes predictions with overall lower performance, with values $r_\phi \geq 0.55$ and $R^2 \leq 0.97$ for \recunet{}'s predictions.

We can observe that the images with $\overline{x}_\mathrm{HI,true} < 0.1$ (blue colour) are spread across the correlation plot, loosely following the dashed black line. This indicates that for images with almost no 21-cm signal (almost completely ionized), the binary prior does not provide additional information and has no evident effect in enhancing the recovery of the 21-cm signal. In these cases, because the neutral regions are few in number and small in size, any error in the binary maps has a sizeable negative impact on \recunet{}'s predictions.

On the other hand, binary priors representing highly neutral scenarios ($\overline{x}_\mathrm{HI,true} \geq 0.9$) are mainly located on the right-most or bottom-right part of the correlation plot. This indicates that binary priors that are almost entirely neutral can mislead \recunet{} predictions even if \segunet{} correctly predict them ($r_\phi \sim 1$). This is explained by the fact that the prior is a simple binary labelling of neutral/ionized regions. In this case, the binary prior is almost entirely uniform, which misleads the predictions as it does not distinguish between the residual foreground and the 21-cm signal. This leads to poor recovery of the 21-cm images ($R^2 \leq 0.4$) for data with $\overline{x}_\mathrm{HI,true} \geq 0.9$.

\begin{table}
    \centering
    \caption{Redshift-bin averaged accuracy and standard deviation for the recovered 21-cm image ($R^2$) and the binary prediction score ($r_\phi$) of \serenet{} at reionization milestones.}
    \begin{tabular}{|M{1.6cm}|M{1.6cm}|M{1.6cm}|} \hline
        $\overline{x}_\mathrm{HI,\,true}$ & $R^2$ $[\%]$ & $r_\phi$ $[\%]$ \\ \hline\hline
        $0.0 - 0.1$ & $81.6 \pm 16.2$ & $63.3 \pm 15.6$ \\\hline
        $0.1 - 0.2$ & $88.5 \pm 3.2$ & $72.5 \pm 4.7$ \\\hline
        $0.2 - 0.3$ & $88.0 \pm 3.0$ & $75.2 \pm 3.7$ \\\hline
        $0.3 - 0.4$ & $87.0 \pm 3.0$ & $77.6 \pm 3.2$ \\\hline
        $0.4 - 0.5$ & $85.4 \pm 3.0$ & $79.7 \pm 2.8$ \\\hline
        $0.5 - 0.6$ & $83.7 \pm 3.3$ & $82.1 \pm 2.5$ \\\hline
        $0.6 - 0.7$ & $82.0 \pm 3.4$ & $85.1 \pm 2.1$ \\\hline
        $0.7 - 0.8$ & $79.0 \pm 3.9$ & $88.3 \pm 1.8$ \\\hline
        $0.8 - 0.9$ & $73.2 \pm 5.3$ & $92.1 \pm 1.7$ \\\hline
        $0.9 - 1.0$ & $50.1 \pm 17.2$ & $97.2 \pm 1.8$ \\\hline
    \end{tabular}\label{tab:accuracy}
\end{table}
\begin{figure*}
	\includegraphics[width=\textwidth]{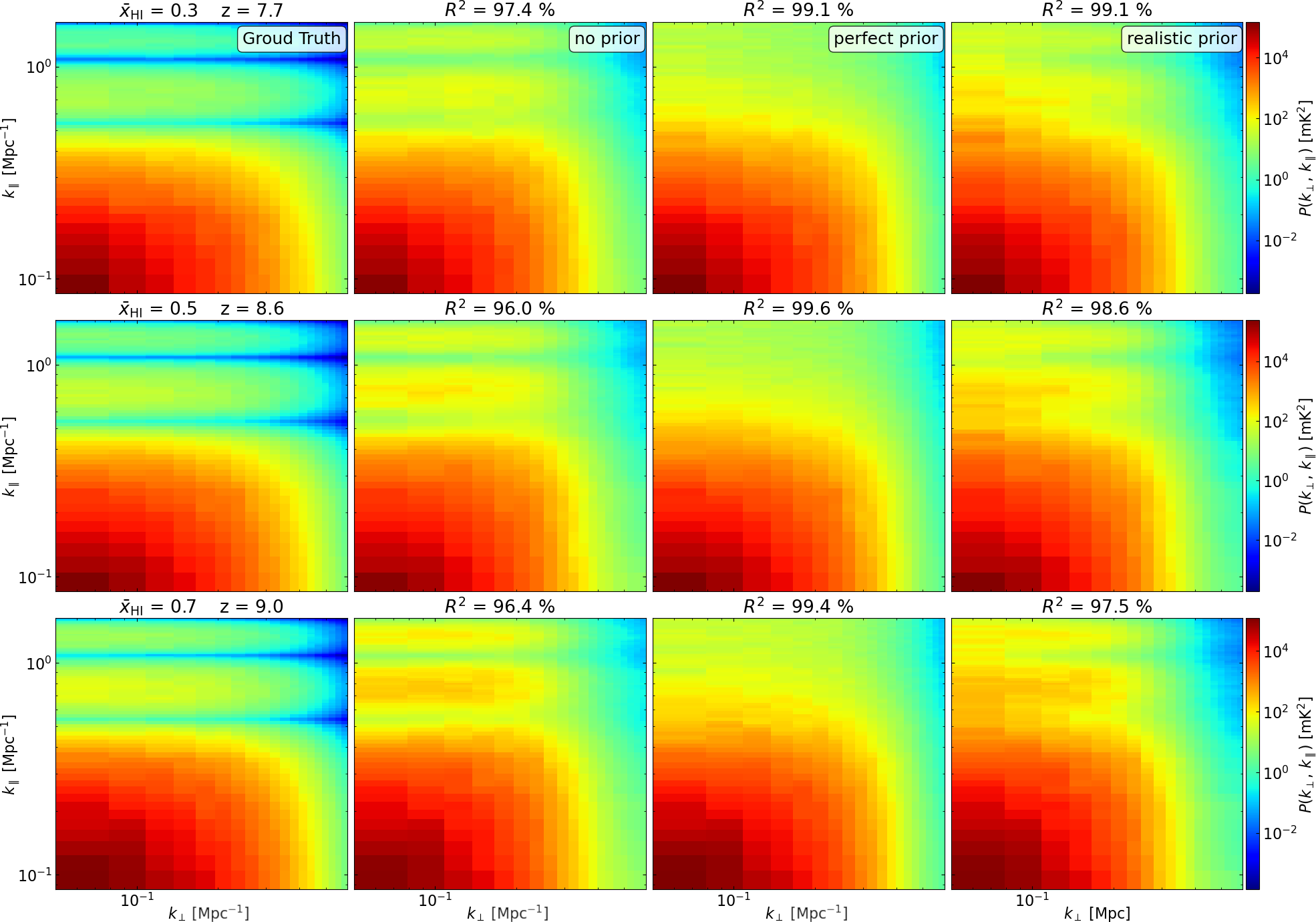}\vskip-2mm
	\caption{Comparison of the predicted cylindrical power spectra for different setups at various central redshifts. From top to bottom, the redshifts are $z = 7.689$, $8.631$, and $9.0$, corresponding to reionization milestones $\overline{x}_\mathrm{HI,,true} = 0.7$, $0.5$, and $0.3$. From left to right, the columns show the ground truth, followed by predictions from \serenet{} for the \textit{no prior} case, the \textit{perfect prior}, and the \textit{realistic prior} case. The results indicate that \serenet{} reproduces the power spectra effectively, with further improvements observed when we include prior information.}
	\label{fig:ps_comparision}
\end{figure*}

\subsection{Recovered cylindrical Power Spectra}\label{sec:recover_ps}
Here, we study the accuracy of predicting the cylindrical (or 2D) power spectra of the \textit{fiducial} model. We use a sub-volume with a frequency width of $20\mathrm{MHz}$, centred at redshifts $z_c = 7.69$, $8.63$, and $9.02$. For the \textit{fiducial} model, these redshifts correspond to the same reionization milestones as shown in \autoref{fig:visual_comparison}. 

\begin{figure*}
	\includegraphics[width=\textwidth]{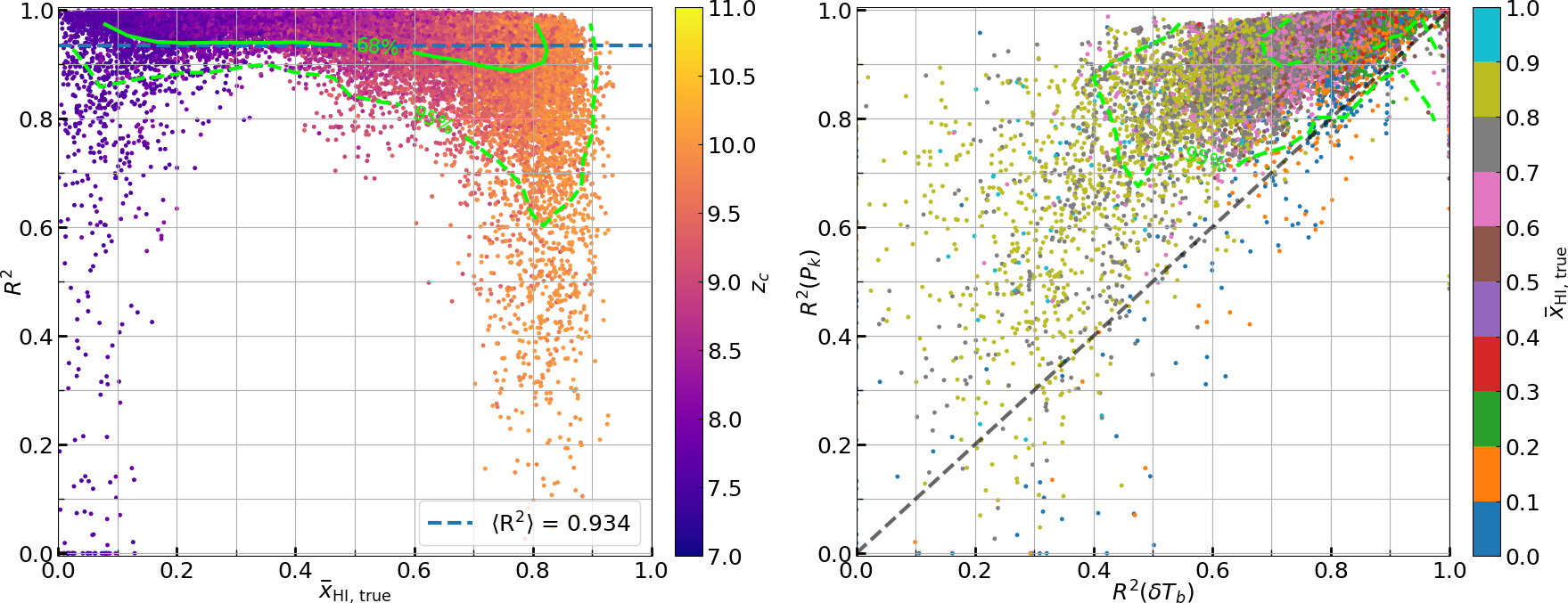}\vskip-2mm
	\caption{\textit{Left panel}: Similar analysis for $R^2$ versus volume-averaged neutral fraction, but the score is calculated on 2D power spectra for the \textit{random dataset}. \textit{Right panel}: comparison of the score computed in the Fourier space, $R^2(P_{k_\perp k_\parallel})$, and the spatial domain $R^2(\delta T_b)$. We randomly select a $100$ redshift within the available range, $7<z<11$, with a frequency depth of $\pm 20\rm\,MHz$. The colorbar indicates the central redshift, $z_c$. The extremes of the redshift range are not sampled due to the required frequency depth for calculating the power spectra.}
	\label{fig:ps_stats}
\end{figure*}
In \autoref{fig:ps_comparision}, we show the cylindrical power spectra of the fiducial model for the three different scenarios. We notice that the accuracy of the predictions is considerably high, with more than $95$ per cent accuracy in all cases. For instance, the \textit{no prior} case also achieves high accuracy despite the recovered 21-cm signal in the spatial domain limited to $R^2<0.85$. As we expected from the results in \S\ref{sec:eff_prior}, the \textit{perfect prior} provides the best result with the astonishing score within $R^2\geq 0.99$ across all redshifts, including the binary maps predicted by \segunet{} as prior information still provides better results than the \textit{no prior} but of only a few per cent. 

\begin{figure*}
	\includegraphics[width=\textwidth]{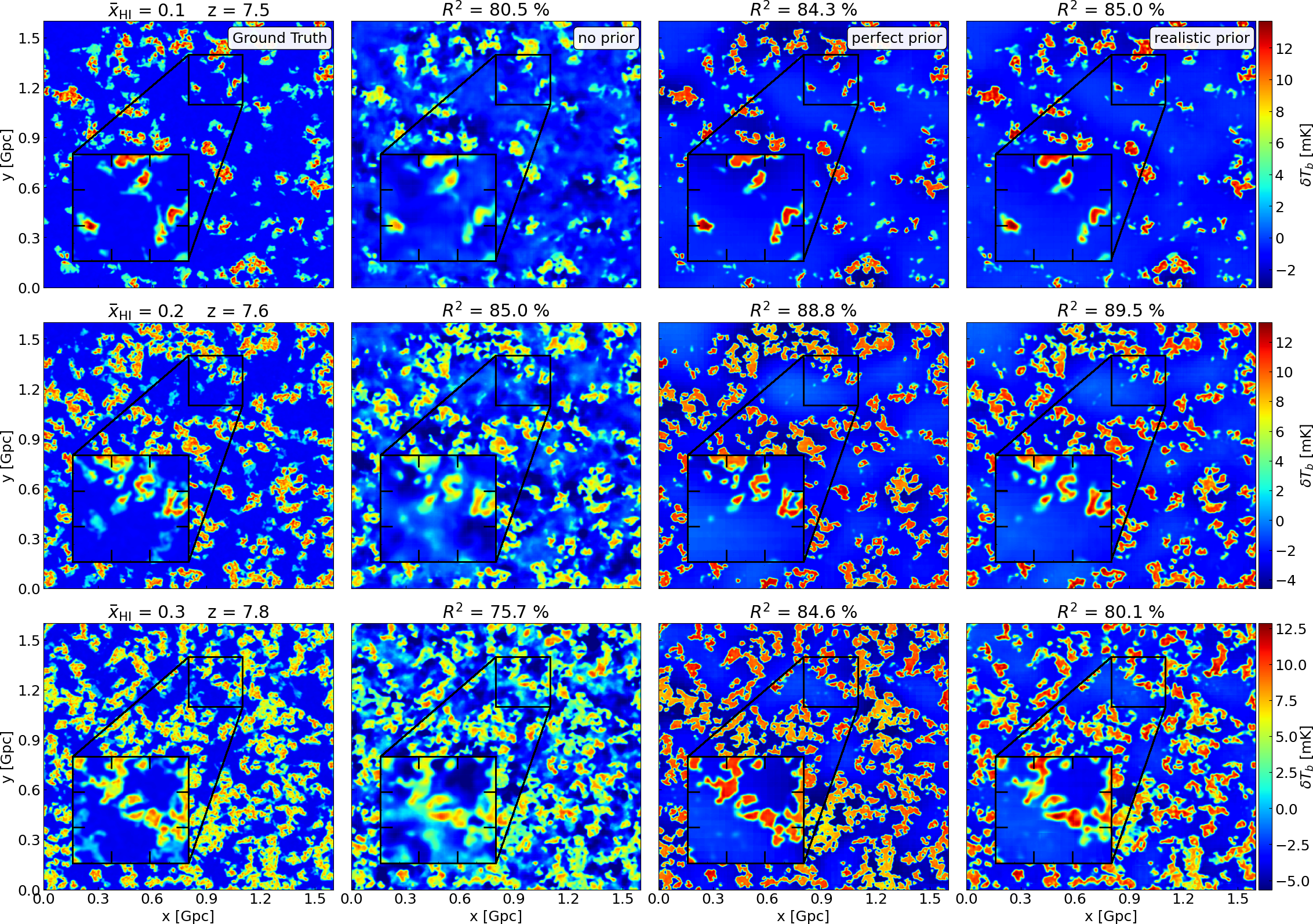}\vskip-2mm
	\caption{Similar visualisation as in \autoref{fig:visual_comparison} comparing different training scenarios applied to the \eos{} data for different reionization milestones mentioned in the panel title in the leftmost column. In each plot, we show an inset plot that zooms in on a region of size 256 Mpc per side, corresponding to the dimensions of the network input.}
	\label{fig:visual_eos}
\end{figure*}
The same features at $k_\parallel \approx 0.55$, $1.05$ and $1.60 \,\rm Mpc^{-1}$ are present in the case of the \textit{no prior} and \textit{realistic prior}, whereas they are less evident in the third case. These features are spectral leakage due to sharp boundaries in the data when computing the 2D power spectrum \citep[e.g.,][]{Barry2019improvMWA, Wilensky2022expl}. This sharp discontinuity is stronger along the $k_\parallel$ direction due to evolution of the 21-cm signal along the line-of-sight, namely the light-cone effect \citep[e.g.,][]{Datta2012Light-coneSpectrum, Giri2018BubbleTomography}. In the \textit{perfect prior}, the recovered 21-cm signal appears to be smoothed and presents fewer fluctuations within the 21-cm regions. Moreover, the large-scale mode, at small $k_\parallel$ and $k_\perp$, appears to be the predominant factor influencing the score estimation, $R^2$. The amplitude of the signal in the 2D power spectra is sensitive to the global average neutral fraction at the corresponding central redshift, $P_\mathrm{21cm}(k_\perp,k_\parallel) \propto T^2_0\, \overline{x}^2_\mathrm{HI}$ \citep{Lidz2007HigherSpectrum, Georgiev2022large}. Therefore, in our case, although the $R^2$ score is not capturing the spectral leakage effect, it indicates how well the neural network captures the reionization history.

An important conclusion is that an almost perfect reconstruction of the power spectra, $R^2 \geq 0.95$, does not imply a recovery of the 21-cm signal of the same degree of accuracy, as is the case for the \textit{no prior} test. Thus, in future observations, we can expect that techniques that aim to mitigate foreground contamination exclusively in the visibility space require additional steps in the image domain to recover the corresponding 21-cm images for astrophysical and cosmological inferences. 

We perform a similar analysis, as shown in \autoref{fig:stat_dataset}, but we apply the score to the 2D power spectra for the $300$ realization of the \textit{random dataset} for the \textit{realistic prior} case. For each realization of the \textit{random dataset}, we randomly choose $100$ redshifts as central redshift, $z_c$, and we then select a $\pm 10\rm MHz$ region around this redshift, from which we calculate $P(k_\perp,\,k_\parallel)$. The results in the left panel of \autoref{fig:ps_stats} confirm that a high score with $R^2 > 0.90$ is achieved by $68$ per cent of the data, while the $95$ per cent contours show results within $R^2 > 0.60$. Similarly to the spatial domain case shown in \S\ref{sec:stat_eval}, the accuracy decreases for samples close to the redshift limits. Overall, we can expect \serenet{} to perform on the same level of accuracy as the average score for the entire \textit{random dataset}, which is $\left< R^2\right> = 0.934$. In this analysis, we have fewer or no data close to the redshift limit, which limits the data to reionization history with $\overline{x}_{\ rmHI,\ true}\leq 0.95$. This is due to the fact that the power spectra are calculated for a sub-sample of the entire lightcone, centred at a given redshift and with a frequency depth.

On the right panel of \autoref{fig:ps_stats}, we show the relation between the score calculated on the predicted power spectra (y-axis), $R^2(P_{k_\perp k_\parallel})$, against the same score computed on the same predicted differential brightness sub-volume (x-axis), $R^2(\delta T_b)$, used to calculate the summary statistic. This analysis calculates the $R^2(\delta T_b)$ on the entire sub-volume, contrary to the same score applied to the images in \autoref{fig:stat_dataset}. The quantity $R^2(P_{k_\perp k_\parallel})$ corresponds to the same quantity on the y-axis as in the left panel. 

From this analysis, we prefer that the accuracy in the spatial and (Fourier) frequency domains will have a linear relation (black dashed line), indicating that the score of the summary statistic reflects the one indicated in the real space. However, as anticipated above, the results are on the left of the one-to-one relation, which indicates that the summary statistic achieves better scores than the real space image. This is perhaps related to the summary statistic and the non-Gaussian nature of the reionization process \citep[e.g.][]{Giri2019Position-dependentReionization}, as more complex features are present in the spatial domain but are not sampled by the power spectra analysis, thus misleading to more accurate prediction.

\subsection{Recover 21-cm signal on Large Field of View}\label{sec:eos_dataset}
In this section, we applied our network to data with a relatively small field of view (FoV) compared to the expected scales that the SKA-Low will provide. In the case of SKA-Low, we expect tomographic images with an angular resolution of $16\,\rm arcsec$ and FoV of $10\,\rm deg$. For reference, in our simulations, at $z=7$, we have a FoV of $\sim1.7\,\rm deg$ and an angular resolution of $\sim46.8\,\rm arcsec$. Therefore, here we test the ability of our network to work on different angular resolutions as well as larger FoV. 

For this purpose, we employ the \textit{Evolution of 21-cm Structure} (\eos{}) project data \citep{Mesinger2016EOS}. This project ran the \texttt{21cmFAST} semi-numerical code, introduced in \S\ref{sec:21cmFast}, in simulation volumes with box size of $1.6\,\rm Gpc$ and a mesh size of $1024$ per side, corresponding to a $\rm{FoV}=10.4\,\rm deg$ at $z=7$. The \eos{} project proposes two galaxy models; here, we employ the faint galaxies model simulations, as this model conforms to our assumption that the spin temperature, $T_\mathrm{S}$, is mostly saturated at high redshift, $z\sim 10$. However, the simulation spin temperature is almost completely saturated at approximately $z\approx 9$. We apply the different steps to the \eos{} data to create a SKA-Low mock observation, see \S\ref{sec:obs_effect}, and, following the discussion in \S\ref{sec:pipeline}, we apply PCA mitigation method to remove most foreground contamination. Because of the large field, we apply the PCA on a sub-volume rather than the entire tomographic dataset. Moreover, due to the different assumption on the $T_\mathrm{S}$ and its implication on the \segunet{} predictions, we are limited to the redshift range $7.0 \leq z \leq 8.2$.

Similar to \S\ref{sec:visual_eval}, we employ \serenet{} to the \eos{} data for the three proposed scenarios. As illustrated in \autoref{fig:network_architecture}, our network works on inputs with a mesh of $128$ per side. Therefore, to apply it to the image of \eos{}, at each redshift, we break up the original $1024^2$ into smaller images that fit the \serenet{} input shape and recombine them together after being processed by our network. We explain our approach in detail in \S\ref{ap:application}.

In \autoref{fig:visual_eos}, we visually compare the recovered 21-cm signal for three scenarios, similarly to \S\ref{sec:visual_eval}. Each panel shows a zoom-in region with the same mesh size as the network input. As we can notice, the \textit{realistic prior} appears to be the best-performing realisation, surpassing the \textit{perfect prior} by several per cent. In both cases, the highest score is achieved toward the central part of the redshift range $z\sim7.6$, when $\overline{x}_\mathrm{HI,true}=0.1$, with values close to $R^2\simeq0.90$. The \textit{no prior} result shows similar results, although less accurate. For instance, we can see from the zoom-in of the panel at $z=7.8$ that the \textit{no prior} case is mistaking detecting a region of 21-cm signal in the lower middle part with $\delta T_b\sim4\,\rm mK$, where in reality there is no \hi{} region. In this case, the \textit{no prior} setup can mistakenly detect some residual foreground contamination as a signal. Instead, the other two scenarios avoid this error thanks to the prior. The \textit{realistic prior} shows constant regions of 21-cm emission with almost no signal fluctuations within the \hi{} region, suggesting that the network relies mainly on the binary prior rather than the residual image, $I_\mathrm{res}$, for the recovery of the 21-cm signal. Meanwhile, the \textit{realistic prior} visually provides a more convincing prediction, showing fluctuations within the region of 21-cm emission shown in the ground truth. We refer the reader to \S\ref{sec:galaxy} for a more thorough discussion on the importance of the prior choice. 
\begin{figure}
	\includegraphics[width=\columnwidth]{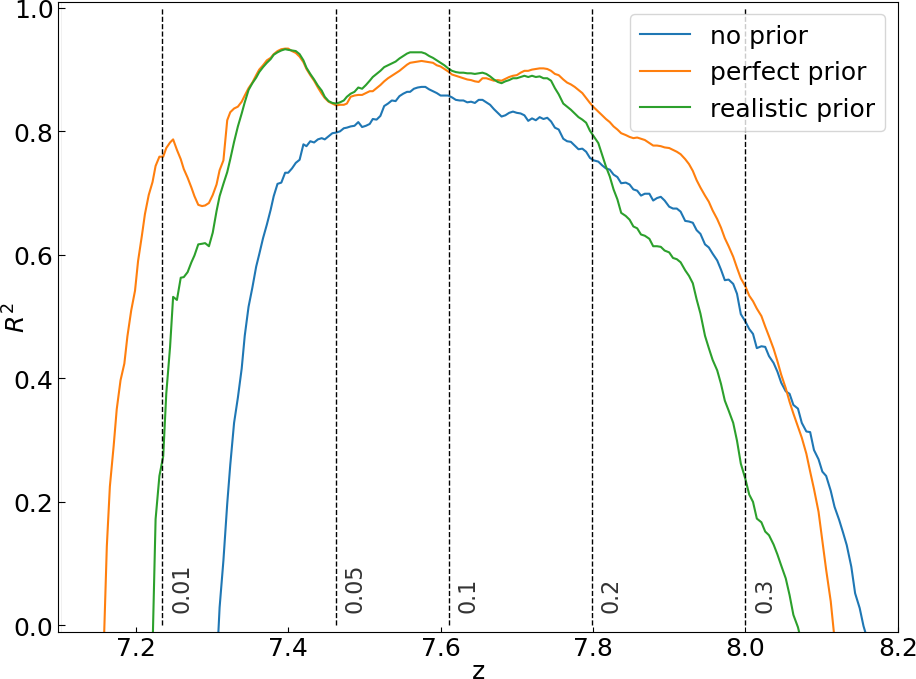}\vskip-2mm
	\caption{Redshift evolution of the $R^2$ score for the prediction by \segunet{} in the \eos{} model, which has a 10-degree field of view larger than that of the training set. Vertical lines indicate the volume-averaged neutral fraction $\bar{x}_\mathrm{HI}$ corresponding to key reionization epochs.}
	\label{fig:stat_eos}
\end{figure}
\begin{figure*}
	\includegraphics[width=\textwidth]{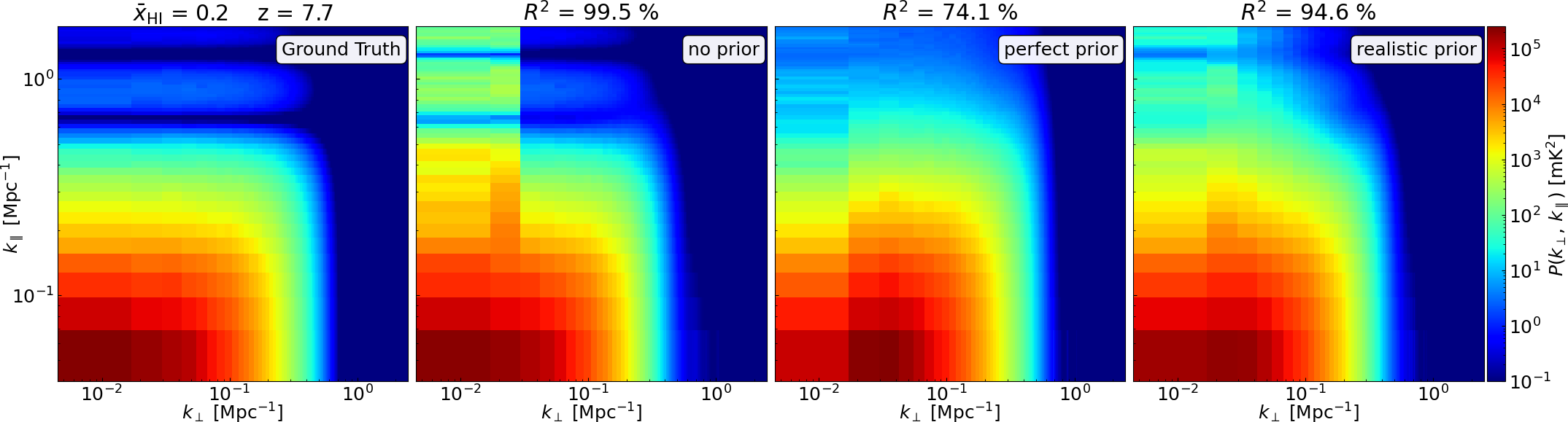}\vskip-2mm
	\caption{Cylindrical power spectra constructed from the \serenet{} predictions of the foreground contaminated \eos{} data at $z=7.7$. Each column is the same as in \autoref{fig:ps_comparision}.}
    \label{fig:ps_eos}
\end{figure*}

In \autoref{fig:stat_eos}, we confirm that the two scenarios with prior have predictions with similar and, at the same time, higher accuracy than the \textit{no prior} results with a difference of $5-10\%$. The redshift evolution of the $R^2$ score shows the characteristic depression at the redshift limits of the subvolume due to the pre-processing. The \textit{perfect prior} shows a wider redshift range of efficiency with lower and upper limits of $z\geq7.15$ and $z\leq 8.1$. Surprisingly, the \textit{no prior} has the same upper limit at the cost of accuracy drops at higher redshift, $z\sim7.35$. Moreover, the accuracy of the \textit{realistic prior} decreases more quickly than the other two cases, so by $z\sim7.8$, the \textit{no prior} provides better prediction. The reason could be related to the fact that at this redshift the simulation still presents a few cold, $T_\mathrm{S}<T_\mathrm{CMB}$, ionized regions that could confuse the segmentation step, as \segunet{} is trained on data that assumes that only regions with no signal, $\delta T_b = 0\,\rm mK$, are indeed ionized $x_\mathrm{HI} = 0$. This misleading does not affect the \textit{perfect prior}, in contrast to the \textit{realistic prior}, as it does not require \segunet{} to infer the binary prior from the residual image. On the other hand, the drop in accuracy at higher redshift for the \textit{no prior} case suggests that, in this case, without the help of the prior, the network mistakes foreground residual as a 21-cm signal.

For completeness, we calculate the 2D power spectra for the \eos{} data and show the result in \autoref{fig:ps_eos}. As in the previous section, \S\ref{sec:recover_ps}, we select the central redshift to be at the centre of the subvolume, where all the models have the highest accuracy, with a $20\,\rm MHz$ frequency depth and compute the cylindrical power spectrum on this subvolume. From the results, we notice that the \textit{no prior} gives much higher accuracy on the power spectra prediction, although it presents an anomalous feature at $k_\perp\leq 0.02\,\rm Mpc^{-1}$ for all $k_\parallel$. Meanwhile, the other scenario visually provides reasonable recovered 2D power spectra. The \textit{realistic prior} is somewhat lower with $R^=0.95$ compared to the other two cases. Meanwhile, the \textit{perfect prior} has a much lower accuracy of $R^2=0.74$, positioning this result as one of the few realisations where $R^2(\delta P_{k_\perp k_\parallel}) < R^2(\delta T_b)$ as discussed in \autoref{fig:ps_stats}. We recognise that this difference is due mainly to the limitation of the $R^2$ metric. The value at $k_\perp=0.01\,\rm Mpc^{-1}$ and $k_\parallel = 0.05\,\rm Mpc^{-1}$ is two orders of magnitude larger than the anomalous features in the \textit{no prior} result. Therefore, the resulting accuracy calculated with \autoref{eq:r2score} is biased to this large contribution rather than a global evaluation at all k-mode.
\subsection{Galaxy Surveys as the Prior Information}\label{sec:galaxy}
In the previous sections, we established the advantage of \serenet{} in using a binary map, derived from the \hi{} distribution, as prior and how this setup is preferable to using a simple U-shaped network. However, in this section, we want to take a step forward, and we want to test if any prior in \serenet{} is better than no prior at all. For this test, we rely on the fact that future and ongoing early Universe galaxy surveys (e.g., JWST, Euclid and Nancy Grace Roman Space Telescope) will eventually provide an exhaustive catalogue with position and redshift of the ionizing sources during the EoR for redshift range within $7\leq z \leq 11$ that are brighter than UV absolute magnitude $M_\mathrm{UV} \approx -12$.

We intend to understand if the proposed architecture (illustrated in \autoref{fig:network_architecture}) could work on a prior based on the angular position and the redshift of these ionizing sources. However, this approach requires some understanding of the source properties and their effect on the surrounding IGM. However, in this test, we assume we know the mean-free path of UV photons at high redshift and consider a redshift-dependent relation with two free parameters:
\begin{equation}\label{eq:mfp}
    \lambda_\mathrm{mfp}(z) = \frac{c}{H(z)}\,\left(\frac{1+z}{z_0}\right)^{\alpha_0} \ . 
\end{equation}
Here, we consider the values of the free parameters $\alpha_0=-2.55$ and $z_0=2.47$ \citep{Choudhury2009analytical}. To be able to apply this method, we employed the publicly available version of \texttt{21cmFAST\_v4}\footnote{\url{https://github.com/21cmfast/21cmFAST/tree/v4-prep}} (Davis et al., in prep.). Contrary to its previous publication, this version can generate halo catalogues identified from the Lagrangian initial conditions using an ellipsoidal collapsed barrier \citep{Sheth2002exc} and then move them to real space using 2LPT.
\begin{figure}
    \includegraphics[width=0.98\columnwidth]{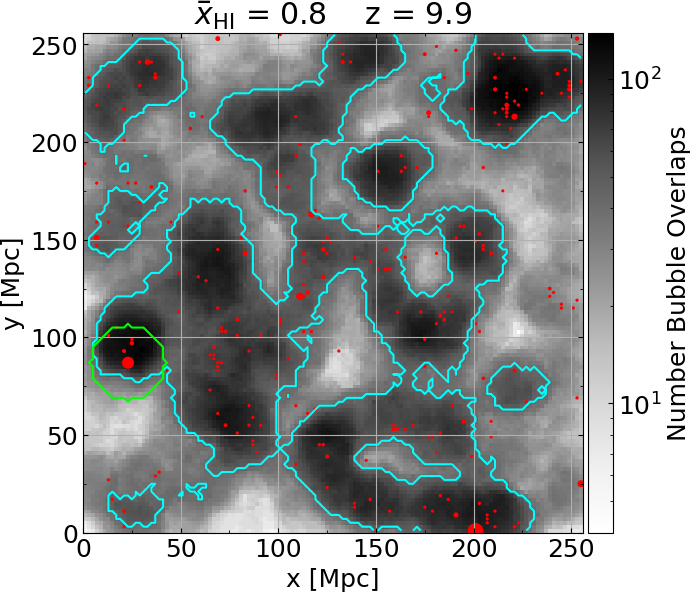}\vskip-3mm
	\caption{An example of the prior created from the source location and the redshift-dependent MFP of UV-photons. The colorbar indicates the number of bubble overlaps. Red dots are the sources, and their size is a reference for the halo mass. Cyan contours indicate the binary maps extrapolated by the bubble overlap map. For example, for the most prominent source, we show the size of the $R_\mathrm{mfp}$ in light green.}
	\label{fig:mfp_prior}
\end{figure}

\begin{figure*}
	\includegraphics[width=\textwidth]{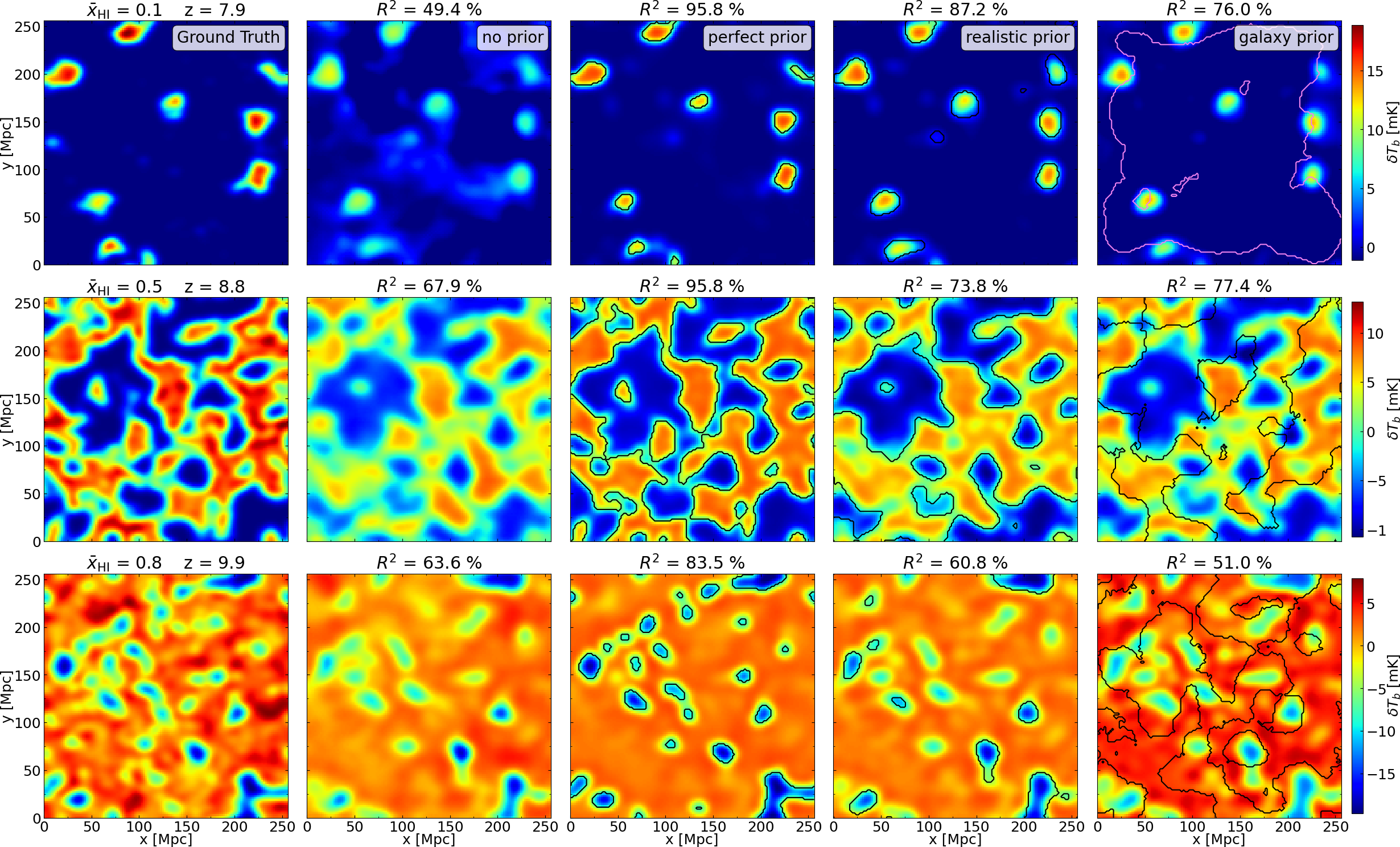}\vskip-2mm
	\caption{Visual comparison for different setups applied to the simulation produced with the version of \texttt{21cmFAST\_v4} that generates halo catalogues. Similar to \autoref{fig:visual_comparison} and \autoref{fig:visual_eos} but with an additional column (right-most panels), showing the \serenet{} results on the \textit{galaxy prior}.}
	\label{fig:visual_mpf}
\end{figure*}

To paint an approximate bubble \hii{} bubble around sources, we follow the approach proposed by \citet{Korber2023pinion}. Based on the position of the sources and their redshift, we apply \autoref{eq:mfp} and define a spherical region around each source, and we expect the size of the \hii{} bubble to be within this distance from the source. In \autoref{fig:mfp_prior}, we show an example of the galaxy prior. The colour map indicates the combined overlap for all sources within the coeval cube, while the red dots correspond to the halo. The size of the red circles is indicative of the halo mass. In green, we show an example of the spherical region described by \autoref{eq:mfp} for the most significant source in the simulation. We create a binary prior from the data with overlapping spheres by taking the average value as a threshold and labelling pixels with value $0$ when the number of overlaps is above the average instead of $1$ if it is below. We refer to this new prior as the \textit{galaxy prior}. We show the corresponding example of the constructed prior map at redshift $z=9.9$ in \autoref{fig:mfp_prior} with cyan-coloured contours. In this case, the threshold is of $\sim 42$ overlaps.

We employ the resulting \textit{galaxy prior} maps as the prior input for \serenet{} as an alternative case of the \textit{realistic prior}. Therefore, without retraining the network, we show the resulting prediction of the 21-cm signal in \autoref{fig:visual_mpf} at redshift $z=9.9$, $8.8$ and $7.9$ corresponding to $\overline{x}_\mathrm{HI,true}=0.8$, $0.5$ and $0.1$. The last column shows the results from the \textit{galaxy prior}. On each panel that uses a prior, we overplot the contour to show the binary prior map employed by the network. 

Here, the \textit{no prior} has a maximum average score of $R^2=0.55$ around redshift $z\approx8$ to then decrease around $R^2=0.47$ for $z\approx9$ and higher redshift. On the other hand, as illustrated in \S\ref{sec:visual_eval} and \ref{sec:eff_prior}, the \textit{perfect prior} provides the best results with an almost constant accuracy of $R^2\sim0.94$ for $z \lesssim 9$ and $R^2=0.81$ at higher redshift. It also shows similar features in the predicted 21-cm images, with virtually no signal fluctuations within \hi{} regions. Moreover, once again, the \textit{realistic prior} provides better results than the \textit{no prior} case with almost up to twice more accuracy at $z=7.9$, $R^2=0.87$, to the same level of accuracy, $R^2\sim 0.60$, at higher redshift, $z\gtrsim 9.5$, as discovered in the previous analysis with \autoref{fig:stat_fiducial}. 

Finally, the newly introduced prior also provides better results than the simple \textit{no prior} case even though the prior in this scenario is far from flawless compared to the other two cases, as shown by the contours in \autoref{fig:visual_mpf}. The \textit{galaxy prior} has an accuracy of $R^2\sim.75$ compared to the $R^2\sim.50$ of the \textit{no prior} case at $z \approx 8$ and similarly to the previous case, its accuracy decreases at higher redshift to an average of $R^2\sim0.50$. Moreover, as expected, the \textit{galaxy prior} performs slightly worse than the  \textit{realistic prior}, suggesting that retraining the network on the new prior with a more appropriate $\lambda_\mathrm{mfp}$ could vastly improve its performance.

Our vital finding is that, counterintuitively, our network trained with a more realistic prior provides more stable predictions. This behaviour is related to the fact that using priors with "imperfections" during the training process makes the network learn not to rely excessively on the binary prior itself, thus allowing the \serenet{} framework to accept prior maps generated from different assumptions, as in the case of the \textit{galaxy prior}. We prove this conclusion because in both cases, the \textit{realistic} and \textit{galaxy prior} can predict fluctuations within the 21-cm signal even though the prior does not provide helpful information in those regions. For example, in the \autoref{fig:visual_mpf}, for the two cases mentioned above at $z=9.9$, we can see that the regions with signal $\delta T_b \approx -5\,\rm mK$ located at $(x,y)\approx(175,225)\,\rm Mpc$ and around $(x,y)\approx(100,175)\,\rm Mpc$ are correctly recovered without the suggestion of the corresponding prior.

With this approach, we aimed to demonstrate that the \serenet{} framework facilitates synergistic studies between SKA-Low 21-cm tomographic datasets and galaxy survey observations. In this context, prior information can enhance the prediction of 21-cm images. We are confident that \serenet{} could also be applied to line-intensity mapping experiments (e.g., [C\scalebox{1.0}[0.9]{II}], [CO]) for cross-correlation with galaxy surveys \citep[e.g.][]{Comaschi2016, Breysse2019}.

\section{Conclusion}\label{sec:conclusion}
In this work, we build upon our previous studies \citepalias{Bianco2021segunet, Bianco2023deep2} by extending our deep learning approach for identifying \hi{} to now recover the 21-cm signal from tomographic images contaminated by synchrotron galactic foregrounds. While this study focuses on SKA-Low radio interferometry observations, the approach applies to any telescope capable of producing 21-cm signal images. Here, we have built upon our previously developed neural network for segmentation, \segunet{}, and introduced a second network, \recunet{}, for recovering the 21-cm signal. 

Diverging from the conventional use of standard U-shaped neural networks in cosmological applications, we have designed a novel architecture for \recunet{}. This architecture incorporates additional input as prior information, a concept inspired by approaches from other fields \citep{Tofighi2019prior, Jurdi2020BBUnet}, to improve signal recovery.
In \S\ref{sec:network_architecture}, we introduce this new deep learning architecture, which modifies the standard U-shaped neural network by allowing a second input, specifically a binary prior. An \textit{Interception Convolution Block} is integrated with the skip connections of the U-Net to process this additional input, thereby incorporating prior information during network training.

We present our key findings in \S\ref{sec:results} by considering three scenarios to test \serenet{}. In the first case, we train a standard U-shape network (\textit{no prior}) to use as a benchmark against architectures that employ priors. For a second setup, we consider the ideal case where the segmentation process, \segunet{}, provides flawless predictions (\textit{perfect prior}) that are employed as additional input in \recunet{}. In the third scenario, we process the residual image with \segunet{} to produce the binary prior (\textit{realistic prior}). This last scenario is the focus of our analysis as it represents a more appropriate situation for SKA-Low observations, where a residual tomographic image is processed by \serenet{}, first with \segunet{} to determine the prior and subsequently by \recunet{} to recover the 21-cm signal.

At first, in \S\ref{sec:visual_eval}, we show an example of the input and output of \serenet{}. We then focus on the \textit{fiducial} model and visually evaluate the recovered 21-cm tomographic data in \S\ref{sec:eff_prior}. We assess the effectiveness of employing binary prior maps derived from the simulated volume-average neutral fraction maps, $x_\mathrm{HI}(\mathbf{r},z)$. Our first conclusion is that the prior has a tangible effect in removing residual foreground with a differential brightness of $\delta T_b\sim 5\,\rm mK$, which would otherwise be mistakenly detected as 21-cm signals. Moreover, the 21-cm signal recovered with a \textit{realistic prior} visually provides more legitimate results when compared to the \textit{perfect prior}.

In \S\ref{sec:stat_eval}, we statistically quantify the prediction on a large dataset that we employ for testing. Our results show that \serenet{} gives consistently better reconstructed 21-cm images when the binary prior is included as an additional input. In particular, the \textit{realistic prior} shows that an average score of $\left<R^2\right>=0.75$ on the entire \textit{random dataset}, with performances that vary between $R^2 \simeq 0.88$ when $\overline{x}_\mathrm{HI}=0.1$ and $R^2\simeq 0.73$ when $\overline{x}_\mathrm{HI}=0.9$, as shown by the analysis in \autoref{tab:accuracy}. Predictions on the binary maps follow a similar trend with $r_\phi\approx 0.73$ at $\bar{x}_\mathrm{HI}=0.1$ and $r_\phi\approx 0.92$ at $\bar{x}_\mathrm{HI}=0.9$. From the recovered 21-cm tomographic dataset, we calculate the 2D power spectra $P(k_\perp,k_\parallel)$. The analysis in \S\ref{sec:recover_ps} shows an astonishing average score of $\left<R^2\right>=0.93$ on the \textit{random dataset}.

Moreover, in \S\ref{sec:eos_dataset}, we apply \serenet{} to a mock observation with a box size of $1.6\,\rm Gpc$, which has a corresponding field of view of approximately $10\,\rm deg$, consistent with expected SKA-Low observations. We decompose the large images into smaller ones that fit the \serenet{} input shape and compare the abovementioned three scenarios. Our network can recover the 21-cm from large FoV SKA-Low mock observations. Moreover, the \textit{realistic prior} provides better results with higher accuracy than the other cases, indicating that the network trained with the \textit{perfect prior} relies excessively on the binary prior, as the recovered 21-cm maps have almost no fluctuations in the regions emitting signals. Thus, imperfections in the binary prior result in a more balanced training, relying more on the residual image and yielding more stable and accurate predictions. Moreover, we notice that the actual limitation in the redshift range on which we can apply \serenet{} is mainly related to limitations in the efficiency of the preprocessing as well as the spin-saturated approximation, $T_\mathrm{S}\gg T_\mathrm{CMB}$, that we impose on our datasets.

Contrary to our initial intuition, one of our significant discoveries is that training our network on a more realistic prior appears to provide more stable and, in some cases, more accurate predictions. This conclusion was first evident from the results in \S\ref{sec:eos_dataset}, and later it is confirmed in \S\ref{sec:galaxy}, where we do a final test and run an additional 21-cm simulation that also provides a galaxy catalogue. From the location of the sources, their redshift and some assumption on the ionizing photons mean-free path, $\lambda_\mathrm{mfp}$, we can compute a different binary prior on the assumption that reionization is an inside-out process and thus assuming that a region void of signal, $\delta T_b = 0\,\rm mK$ is ionized, i.e. $x_\mathrm{HI}(\mathbf{r},z) = 0$.

This result highlights the potential of using \serenet{} for synergistic observations between SKA-Low and high-redshift galaxy surveys, which could improve predictions and incorporate additional physical information into the network, though astrophysical assumptions may influence this. We also anticipate that \serenet{} could be effectively applied to other line intensity mapping (LIM) experiments at lower redshifts for cross-correlation with galaxy surveys (e.g., {[C\scalebox{1.0}[0.9]{II}]}, [CO]).

There have been several attempts to address foreground residuals in 21-cm observations \citep[e.g.][]{GagnonHartman2021, Kennedy2024}. However, these efforts have not focused on reconstructing the complete 21-cm signal image data, which contains valuable non-Gaussian information. Moreover, similar works have combined standard foreground mitigation techniques with neural network approaches. A similar approach has been applied to Post-EoR \citep{Makinen2021deep, Shi2024}. However, only recently \cite{Chen2024stability} has tested the stability and limitations of these methods to unexpected systematic and different models employed in the training/validation of the network. Our work provides better results than previous works in recovering both \hi{} and \hii{} regions at small scales. It also provides a comprehensive statistical analysis of the recovery efficiency across a wide range of astrophysical parameters and redshifts \citep{Li2019Sep, Sabti2024gen}, indicating that in some scenarios, neural networks encounter greater difficulties in recovering the 21-cm signal. This latter point requires further investigation and will be part of our future work.

With \serenet{}, we can explore higher-order summary statistics, such as the bispectrum \citep[e.g.][]{Majumdar2018QuantifyingBispectrum,Watkinson2019TheHeating} and improve our understanding of the underlying astrophysical and cosmological processes that affect the signal \citep[e.g.][]{jennings2020analysing, hutter202021, giri2022imprints, schaeffer2023beorn, schneider2023cosmological}.

\section*{Acknowledgements}
The authors would like to thank Andrei Mesinger and Adrian Liu for their valuable discussions and comments. The authors would also like to thank the organizers of the \textit{"Cosmic Dawn at High Latitude"} Nordita workshop program for providing a conducive platform to develop this project further. MB acknowledges the financial support from the Swiss National Science Foundation (SNSF) under the Sinergia Astrosignals grant (CRSII5\_193826). We acknowledge access to Piz Daint at the Swiss National Supercomputing Centre, Switzerland, under the SKA's share with the project ID sk014. This work has been done in partnership with the SKACH consortium through SERI funding. Nordita is supported in part by NordForsk.

The deep learning implementation was possible thanks to the application programming interface of \texttt{Tensorflow} \citep{Tensorflow2015} and \texttt{Keras} \citep{Chollet2017}. The algorithms and image processing tools operated on our data were performed with the help of \texttt{NumPy} \citep{numpy2020}, \texttt{SciPy} \citep{scipy2020}, \texttt{scikit-learn} \citep{scikitlearn2011} and \texttt{scikit-image} \citep{scikit2014} packages. All figures were created with \texttt{mathplotlib} \citep{Hunter2007}.

\section*{Data Availability} 
The data underlying this article is available upon request and can also be regenerated from scratch using the publicly available \texttt{21cmFAST} \citep{Mesinger2011,Murray202021cmFAST} and \texttt{Tools21cm} \citep{Giri2020t2c} code. The \serenet{} code and its trained network weights are available on the author's \texttt{GitHub} page: \url{https://github.com/micbia/serenet}.

\bibliographystyle{mnras}
\bibliography{serenet}

\appendix

\section{Application to Large Images}\label{ap:application}
\begin{figure}
	\includegraphics[width=\columnwidth]{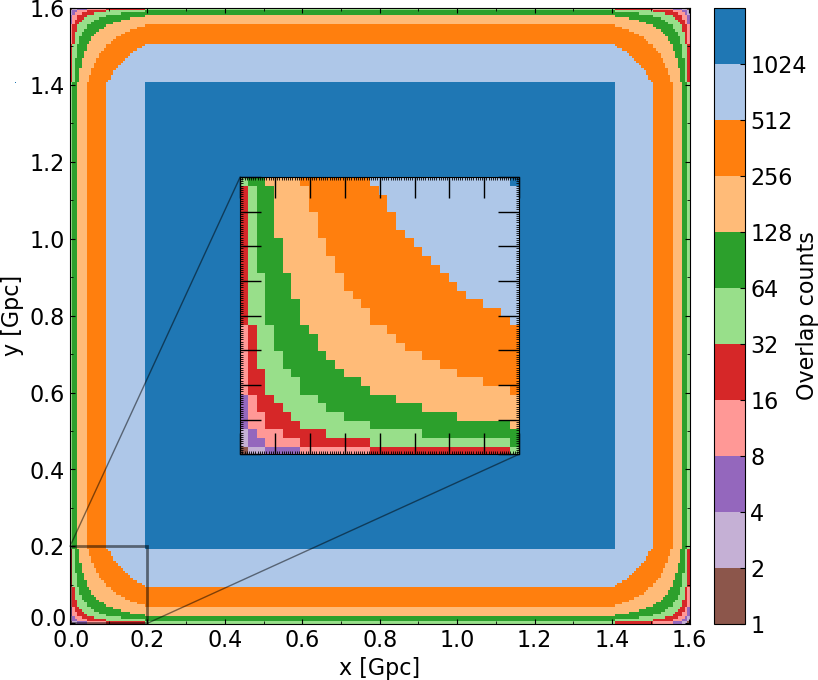}\vskip-2mm
	\caption{Example of \serenet{} applied to an image with a large FoV. The inset plot shows a region with a mesh size of $128^2$ and a comoving size of $1.56\,\rm Mpc$ located at the bottom left corner. This dimension corresponds to the input shape of our networks. The colorbar indicates the number of times the overlapping kernel samples each pixel.}
	\label{fig:largefov}
\end{figure}

Although \serenet{} has been trained on 2D image data with a mesh size of $128$ per side, it can be easily applied to images with larger mesh dimensions. In \autoref{fig:largefov}, we show an example applied to an image with $1024^2$ pixels. We use a kernel with an overlapping factor of $\frac{1}{32}$. This factor indicates the portion of the image that overlaps. For instance, a factor of $1$ is the case where each kernel is applied back-to-back with no overlap, while $\frac{1}{2}$ is the case where half the image is overlapped.

The inset plot shows a zoom-in on the bottom left corner of the larger image. The zoom-in size corresponds to the kernel dimensions and the shape of the \serenet{} input, as mentioned in \S\ref{sec:pipeline}. As we can see from the image, the pixels on the borders of the image are undersampled, and only the very first $4\times4$ pixels are sampled only once. However, from the SKA-Low tomographic image with $\rm{FoV}\simeq10\,\rm deg$, we expect that only the central $4\times4\,\rm deg$ region will be processed for summary statistic analysis.

Because of the selected overlapping factor, this setup virtually creates $\num{50176}$ of partially overlapping images that \serenet{} need to process and eventually puzzle back together. Overlapping pixels are combined with a simple average based on the number of overlaps, indicated by the colorbar in \autoref{fig:largefov}. 

Thanks to the sliding overlapping kernel approach, \serenet{} can be applied to images with much larger mesh sizes, thereby avoiding the need for retraining the network and saving computational time.

\section{The Size of the Dataset}\label{ap:datasize}
\begin{figure}
	\includegraphics[width=1.05\columnwidth]{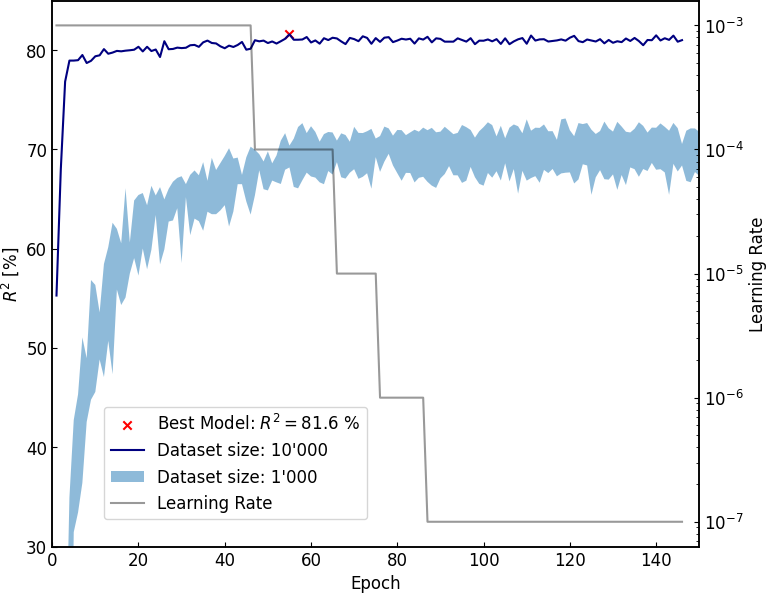}\vskip-2mm
	\caption{Comparison of the $R^2$ score on the validation dataset. The dark blue line indicates the model employed in the results of this paper, which was trained on 10,000 samples. A grey line indicates the learning rate. The blue shadow indicates the corresponding range for models trained on a dataset with one-tenth of the original samples.}
	\label{fig:loss_comp}
\end{figure}

Due to limited computational resources, obtaining a sufficiently large dataset can be challenging. Therefore, we noticed that previous works \citep[e.g.,][]{Gagnonhartman2021recovering, Floss2024} use a significantly smaller dataset when training 3D U-Nets, consisting of a few hundred samples. To the best of our knowledge, only recently has \cite{Sabti2024gen} employed a dataset of comparable size to our previous works \citepalias{Bianco2021segunet, Bianco2023deep2}.

Here, we want to analyse and evaluate the required dataset size for a stable and accurate recovery of the 21-cm signal with U-Nets in foreground-contaminated images. Therefore, we trained a few additional models on a fraction of the entire dataset for the \textit{realistic prior} scenario. For each independent training session, we randomly select 1,000 samples, corresponding to one-tenth of the whole dataset, and train the model until convergence and early stopping is reached.

From \autoref{fig:loss_comp}, we show the $R^2$ score, \autoref{eq:r2score}, on the validation set for these two scenarios. The dark blue line indicates the score for the model trained on the entire dataset, which consists of 10'000 samples, presented in \S\ref{sec:21cmFast} and used for the results of this paper, \S\ref{sec:results}. In this case, we can notice that the score quickly rises after a few epochs and stabilises at $R^2 \simeq 80\%$, reaching its best performance after approximately $55$ epochs. On the other hand, the blue shadow region indicates the corresponding score for the models trained on a reduced dataset. We can notice that for this scenario, the score range is lower, with almost $10\%$ less accuracy. Moreover, the best performance is reached only after $\approx 120$ epochs with an averaged best score around $R^2 \simeq 70\%$. 

The reason for this trend could be related to the fact that the limits of the parameter space we employ, $(\zeta, M_\mathrm{min}, R_\mathrm{mfp})$, are wide, providing a vast range of possible combinations. A similar conclusion was reached in a completely different domain but with similar challenges \citep[e.g.:][]{app11020796, DAWSON2023105284}. Therefore, we conclude that training our network on only a few thousand realisations of the residual foreground contaminated lightcone is insufficient to sample the astrophysical parameter space we employ and generalise the recovery of the 21-cm signal. Moreover, our evaluation suggests that decreasing the range of the astrophysical/cosmological parameter space is necessary to avoid underperformance due to smaller datasets.

\label{lastpage}
\end{document}